\documentclass[12pt]{article}
\usepackage{times}
\usepackage{epsfig}
\usepackage{amsmath}
\usepackage{amsfonts}
\usepackage{color}
 \usepackage{slashed}

\def\beq{\begin{equation}}
\def\eeq{\end{equation}}
\def\bea{\begin{eqnarray}}
\def\eea{\end{eqnarray}}
\def\beqn{\begin{eqnarray}} \def\eeqn{\end{eqnarray}}
\def\beeq{\begin{eqnarray}}
\def\eeeq{\end{eqnarray}}

\def\ep{\epsilon}

\def\nn{\nonumber}
\def\Eq#1{Eq.~(\ref{#1})}
\def\ln#1{\mathrm{log}\left(#1\right)}

\def\li#1{\mathrm{Li_2}\left(#1\right)}
\def\bp{p\hspace{-.42em}/\hspace{-.07em}}
\def\bq{q\hspace{-.42em}/\hspace{-.07em}}

\def\td#1{\tilde{\delta}\left(#1\right)}

\newcommand{\la}{\langle}
\newcommand{\ra}{\rangle}

\def\qb{\mathbf{q}}
\def\pb{\mathbf{p}}

\def\M#1{{\cal M}^{(#1)}} 
 
\def\G#1{\Gamma^{(#1)}}

\def\qq{{q \bar q}}

\newcommand\as{\alpha_{\mathrm{S}}}
\newcommand\g{g_{\mathrm{S}}}

\def\uv{{\rm UV}}
\def\ir{{\rm IR}}
\def\os{{\rm OS}}

\def\cg{c_{\Gamma}}
\def\cgt{\widetilde{c}_{\Gamma}}
\def\gs{g_{\rm S}}
\def\as{\alpha_{\rm S}}
\def\aas{\frac{\as}{4\pi}}
\def\v{{\rm V}}
\def\r{{\rm R}}

\def\re{{\rm Re}}

\def\xiu{\xi_{1,0}}
\def\xid{\xi_{2,0}}
\def\xit{\xi_{3,0}}
\def\xiuv{\xi_{\uv}}

\def\aas{\frac{\as}{4\pi}}

\def\Se{\widetilde{S}_{\ep}}

\begin{document}

\begin{titlepage}
\renewcommand{\thefootnote}{\fnsymbol{footnote}}
\begin{flushright}
     IFIC/16-45
     \end{flushright}
\par \vspace{10mm}
\begin{center}
{\LARGE \bf 
Four-dimensional unsubtraction \vspace{2mm} \\ with massive particles}
\end{center}
\par \vspace{2mm}
\begin{center}
{\bf Germ\'an F. R. Sborlini}~$^{(a)}$\footnote{E-mail: german.sborlini@ific.uv.es},
{\bf F\'elix Driencourt-Mangin}~$^{(a)}$\footnote{E-mail: felix.dm@ific.uv.es} and
{\bf Germ\'an Rodrigo}~$^{(a)}$\footnote{E-mail: german.rodrigo@csic.es}
\vspace{5mm}

${}^{(a)}$ Instituto de F\'{\i}sica Corpuscular, 
Universitat de Val\`{e}ncia -- 
Consejo Superior de Investigaciones Cient\'{\i}ficas, 
Parc Cient\'{\i}fic, E-46980 Paterna, Valencia, Spain

\end{center}

\par \vspace{2mm}
\begin{center} {\large \bf Abstract} \end{center}
\begin{quote}
We extend the four-dimensional unsubtraction method, which is based on the loop-tree duality (LTD), 
to deal with processes involving heavy particles. The method allows to perform the summation over degenerate 
IR configurations directly at integrand level in such a way that NLO corrections can be implemented 
directly in four space-time dimensions. We define a general momentum mapping between the 
real and virtual kinematics that accounts properly for the quasi-collinear configurations, and 
leads to an smooth massless limit. We illustrate the method first with an scalar toy example, and 
then analyse the case of the decay of a scalar or vector boson into a pair of massive quarks. 
The results presented in this paper are suitable for the application of the method to 
any multipartonic process.   
\end{quote}

\par \vspace{5mm}

\vspace*{\fill}
\begin{flushleft}
August 3, 2016
\end{flushleft}

\end{titlepage}

\setcounter{footnote}{0}
\renewcommand{\thefootnote}{\fnsymbol{footnote}}

\section{Introduction}
\label{sec:introduction}
The development of new computational techniques to obtain more accurate theoretical predictions at colliders has been strongly pushed forward by the high precision experimental data obtained from the LHC. In the framework of perturbative quantum field theory (and perturbative QCD in particular), the presence of infrared (IR) and ultraviolet (UV) divergences constitutes the main difficulty that must be overcome to obtain physical results. renormalisation has been proven to deal successfully with the UV structure of these theories, 
and the cancellation of IR singularities is guaranteed by general theorems~\cite{Kinoshita:1962ur, Lee:1964is} for physical observables
that sum over all the possible degenerate IR configurations. This requires taking into account both loop scattering amplitudes and real processes with the emission of additional external particles; the sum of all contributions leads to IR-safe observables. On the other hand, UV divergences are removed by suitable counter-terms, whose divergent structure depends only on the specific theory under consideration.

In order to make these divergences manifest explicitly, the standard approach relies in the introduction of a convenient regularisation method. Nowadays, the standard choice in gauge theories is dimensional-regularisation (DREG) \cite{Bollini:1972ui, 'tHooft:1972fi, Cicuta:1972jf, Ashmore:1972uj} because it preserves gauge invariance. Within DREG, the space-time is analytically continued from $d=4$ to $d=4-2\ep$ dimensions; both UV and IR divergences appear as $\ep$-poles. Using $\ep$ as a regulator, it is possible to perform the loop and the phase-space integrals for the virtual and real-radiation amplitudes, respectively. Thus, the poles of the virtual corrections are cancelled with those present in the real-radiation contributions (due to soft/collinear configurations) and those included in the UV counter-terms. The usual procedure in this framework is to regularise each contribution separately, integrate the expressions and, finally, cancel the $\ep$ poles and take the limit $\ep \to 0$.
 
Besides the renormalisation and the regularisation method, the fact that real and virtual contributions have the same IR-divergent behaviour is the underlying basis of the subtraction methods~\cite{Kunszt:1992tn,Frixione:1995ms,Catani:1996jh,Catani:1996vz}. The general idea of subtraction is the introduction of counter-terms which mimic the local IR behaviour of the real components and that can easily be integrated analytically. In this way, the integrated form is combined with the virtual component whilst the unintegrated counter-term cancels the IR poles originated from the phase-space integration of the real-radiation part. The \emph{subtraction paradigm} has evolved to different versions from the FKS-subtraction~\cite{Kunszt:1992tn,Frixione:1995ms}, and dipole-subtraction~\cite{Catani:1996jh, Catani:1996vz}, to antenna-subtraction~\cite{GehrmannDeRidder:2005cm}, $q_T$-subtraction~\cite{Catani:2007vq,Catani:2009sm} and other recent variations \cite{Czakon:2010td,Bolzoni:2010bt,DelDuca:2015zqa,Boughezal:2015dva,Gaunt:2015pea,DelDuca:2016ily,DelDuca:2016csb}. The treatment of massive particles has also been considered specifically~\cite{Catani:2002hc,GehrmannDeRidder:2009fz,Abelof:2012he,Abelof:2014fza,Bonciani:2015sha}. However, all these approaches treat separately real and virtual corrections, since the final-state phase-space of the different contributions involves different numbers of particles. The construction of IR counter-terms is inherent to the subtraction approach, and it constitutes the main restriction for an efficient application to multi-leg multi-loop processes.

With the purpose of obtaining a fully-local cancellation of IR singularities without building IR-counter-terms, we apply the loop-tree duality (LTD) theorem \cite{Catani:2008xa,Rodrigo:2008fp,Bierenbaum:2010cy,Bierenbaum:2012th} to manage the virtual corrections. This theorem asserts that loop integrals are expressible as the sum of \textit{dual integrals}, which are built starting from tree-level like objects and replacing the loop measure with an extended phase-space measure. The main advantage of LTD is that by introducing additional physical on-shell particles, dual integrals and real-radiation contributions exhibit a similar structure. Moreover,  the loop 
threshold and IR singularities are always restricted to a compact region of the loop three-momentum space~\cite{Buchta:2014dfa,Buchta:2015xda}. These two properties of LTD allow a natural integrand-level combination of virtual and real corrections. 
The method has been recently developed for processes involving massless particles in Refs.~\cite{Hernandez-Pinto:2015ysa,Sborlini:2015uia,Sborlini:2016fcj,Sborlini:2016gbr,ll2016}. The key point in this framework is the establishment of a momentum mapping to generate the real-radiation kinematics from the Born and the loop momenta. In this way, we guarantee not only a simultaneous cancellation of IR singularities without the necessity of introducing IR subtractions but also a fully-local four-dimensional implementation.

Before describing the content of this article, we would like to highlight that many other attempts have been previously developed to obtain four-dimensional representations of higher-order corrections to physical observables. In Refs. \cite{Soper:1998ye, Soper:1999xk, Soper:2001hu, Kramer:2002cd} it was proposed to cancel the singularities through the application of a momentum smearing to relate real and virtual kinematics. In this way, both real and virtual terms could be combined at the integrand level, thus achieving a local cancellation of singularities. Other alternative methods consist in rewriting the standard IR/UV subtraction counter-terms in a local form, as discussed in Refs. \cite{Becker:2010ng, Becker:2012aqa}, or modifying the structure of the propagators and the Feynman rules \cite{Pittau:2012zd,Donati:2013iya,Fazio:2014xea} to regularise the singularities at integrand level. LTD has also been used recently to deal with integral representations of virtual and real subtraction terms~\cite{Seth:2016hmv}, 
including the description of initial-state singularities, that are grouped together and then are integrated numerically.  

The main purpose of this article is to extend the LTD four-dimensional unsubtraction method presented in Refs.\cite{Hernandez-Pinto:2015ysa,Sborlini:2015uia,Sborlini:2016fcj,Sborlini:2016gbr,ll2016} to deal with massive particles. From the kinematical point of view, the mass of the particles slightly modifies the momentum mapping used to perform the real-virtual combination and this changes the IR-divergent structure. 
The major difference comes from the treatment of the self-energy corrections, since they must fulfill non-trivial constraints both in the IR and UV regions. Moreover, quasi-collinear configurations are mapped in such a way that logarithmic contributions are cancelled at the integrand level and the massless limit is smooth.  In any case, the dual representations involve dealing with higher-order poles 
in LTD~\cite{Bierenbaum:2010cy,Bierenbaum:2012th}. 

The outline of this article is the following. In Sec. \ref{sec:reviewLTD} we set the notation and briefly recall some results related with the LTD theorem. After that, in Sec. \ref{sec:massivetriangle}, we apply LTD to study the one-loop scalar three-point function with massive particles. In Sec. \ref{sec:scalarDREG}, we introduce an scalar toy example and compute NLO corrections through the application of the conventional DREG approach. After describing the real emission phase-space partition and introducing a proper momentum mapping in Sec. \ref{sec:realvirtualmap}, we deal with the NLO corrections of the scalar toy model within the LTD approach in Sec. \ref{sec:unsubtraction}. We use LTD and the momentum mapping to perform the real-virtual combination at integrand level and define purely four-dimensional integrable expressions. 
In Sec.~\ref{ssec:renormalisation}, we present integral representations for the wave function and mass renormalisation 
factors for heavy quarks in the on-shell renormalisation scheme. Renormalisation is then discussed in Sec.~\ref{sec:renormalisation}.
Afterwards, we proceed to implement this technique to deal with physical theories. In particular, we compute the NLO QCD corrections to the decay rate $A^* \to q \bar q (g)$, with massive quarks and $A=\{\phi,\gamma,Z\}$. The results are presented in Sec. \ref{sec:physicalexamples}, emphasising the four-dimensional nature of the implementation. Finally, in Sec.\ref{sec:conclusions}, the conclusions are exposed and we discuss future implications of this work.

\section{Loop-tree duality: concepts and notation}
\label{sec:reviewLTD}
\begin{figure}[htb]
\begin{center}
\includegraphics[width=0.3\textwidth]{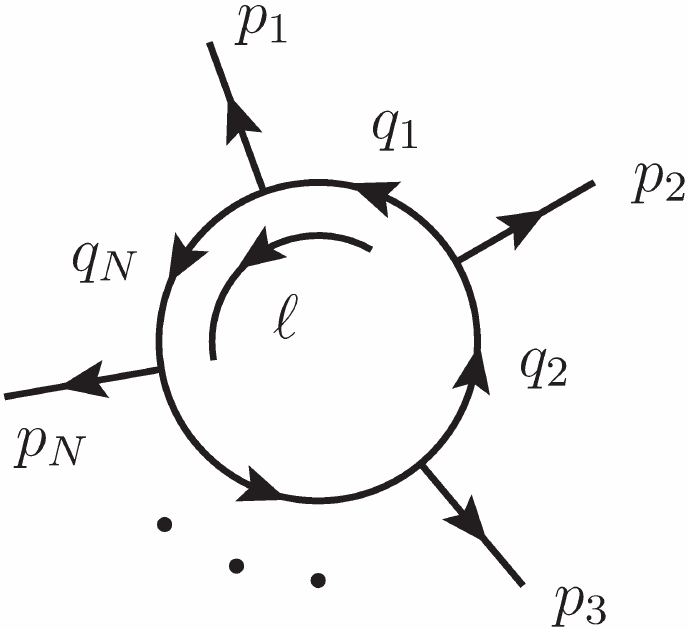}
\caption{One-loop topology with $N$ external legs and the corresponding momenta configuration. 
The external momenta are considered outgoing and the internal momentum flows counter-clockwise.
\label{fig:LTDreview}}
\end{center}
\end{figure}

In this section, we summarise the key concepts of the LTD theorem at one-loop. So, let's consider a generic one-loop scalar integral for an $N$-particle process, as depicted in Fig.~\ref{fig:LTDreview}. If the external momenta are labeled as $p_i$ with $i\in \{1,2,\ldots N\}$, then the internal virtual momenta are given by
\beq
q_{i} = \ell + k_i~, \qquad k_{i} = p_{1} + \ldots + p_{i}~,
\eeq
with $\ell$ the loop internal momentum and $k_N=0$ due to momentum conservation. The corresponding expression for the scalar integral is
\beq
L^{(1)}(p_1, \dots, p_N) = \int_{\ell} \, 
\prod_{i=1}^{N} \,G_F(q_i)~,
\label{eq:oneloop}
\eeq 
where
\beq
G_F(q_i) = \frac{1}{q_i^2-m_i^2+\imath 0} \, ,
\eeq
is the scalar Feynman propagator associated to a virtual particle with mass $m_i$ and four-momentum
$q_{i,\mu} = (q_{i,0},\mathbf{q}_i)$ ($q_{i,0}$ is the energy and $\qb_{i}$ are 
the spatial components). We recall that the $+\imath 0$ prescription is introduced to separate, in the imaginary axis, the solutions arising from the on-shell condition, i.e. $G_F(q_i)^{-1}=0$. In particular, this translates into two solutions with positive and negative imaginary components, respectively. On the other hand,
\beq
\int_{\ell}  
= - \imath \mu^{4-d} \int \frac{d^d \ell}{(2\pi)^{d}} ~,
\eeq
denotes the standard one-loop integration measure in $d$-dimensions.

According to the LTD theorem, any loop contribution to scattering 
amplitudes in any relativistic, local and unitary quantum field theory 
can be computed through dual integrals, which are built from single cuts 
of the virtual diagrams at one-loop~\cite{Catani:2008xa}. 
In other words, there exists a formal connection among loop and phase-space integrals. 
The cut condition is implemented by restricting the integration measure through the introduction of
\beq
\td{q_i} \equiv 2 \pi \, \imath \, \theta(q_{i,0}) \, \delta(q_i^2-m_i^2) \, ,
\label{eq:tddefinition}
\eeq
which forces to integrate in the positive energy mode ($q_{i,0}>0$) of the corresponding on-shell hyperboloid (i.e. $q_i^2=m_i^2$). It is worth appreciating that, in the massless limit, all the hyperboloids associated with the on-shell condition $G_F(q_i)^{-1}=0$ degenerate into light-cones. Considering the one-loop scalar integral, the LTD theorem establishes that its dual representation is given by
\beq
L^{(1)}(p_1, \dots, p_N) = - \sum_{i=1}^N \, \int_{\ell} \, \td{q_i} \, \prod_{j\neq i} \,G_D(q_i;q_j)~, 
\label{eq:oneloopduality}
\eeq 
i.e. the sum of $N$ dual integrals, each one associated with one of the possible single-cuts. In \Eq{eq:oneloopduality}, the dual propagators are
\beq
G_D(q_i;q_j) = \frac{1}{q_j^2 -m_j^2 - \imath 0 \, \eta \cdot k_{ji}} \, ,
\eeq
with $i,j \in \{1,2,\ldots N\}$, $k_{ji} = q_j - q_i$ and $\eta$ an arbitrary future-like or light-like vector, $\eta^2 \ge 0$, with positive definite energy $\eta_0 > 0$ \footnote{The importance of the modified prescription is deeply explored in Ref. \cite{Catani:2008xa}, where the authors emphasise its equivalence with the Feynman-Tree theorem (FTT)~\cite{Feynman:1963ax,Feynman:1972mt}.}. Since $\eta$ is arbitrary, we can chose $\eta^\mu = (1,{\bf 0})$ to simplify the computations. 

Assuming that there are only single powers of the Feynman propagators, the dual representation 
in~\Eq{eq:oneloopduality} is straightforwardly valid for loop scattering amplitudes.
The single-cuts do not affect numerators, therefore, the dual representation of scattering amplitudes is obtained by simply 
adding all possible \emph{dual} single-cuts of the original loop diagram, and replacing the  
uncut Feynman propagators by dual propagators.  If there are higher-powers of the propagators, however,
we should apply the extended version of the LTD theorem~\cite{Bierenbaum:2012th}
 by using the Cauchy's residue theorem with the well-known formula for poles of order $n$, i.e.
\beqn
{\rm Res}({\cal A},q_{i,0}^{(+)}) &=& \frac{1}{(n-1)!} \, \left. \frac{\partial^{n-1}}{\partial^{n-1} \, 
q_{i,0}}\left({\cal A}(q_{i,0})\, (q_{i,0}-q_{i,0}^{(+)})^{n}\right)\right|_{q_{i,0}=q_{i,0}^{(+)}},
\label{eq:RESMULTIPLE}
\eeqn
where $q_{i,0}^{(+)} = \sqrt{\qb_i^2+m_i^2-\imath 0}$ is the positive energy solution of the corresponding on-shell dispersion relation.
In that case, the explicit form of the scattering amplitude is relevant because the numerator is affected by the derivative. 

\section{Massive scalar three-point function within LTD}
\label{sec:massivetriangle}
\begin{figure}[t]
\begin{center}
\includegraphics[width=0.4\textwidth]{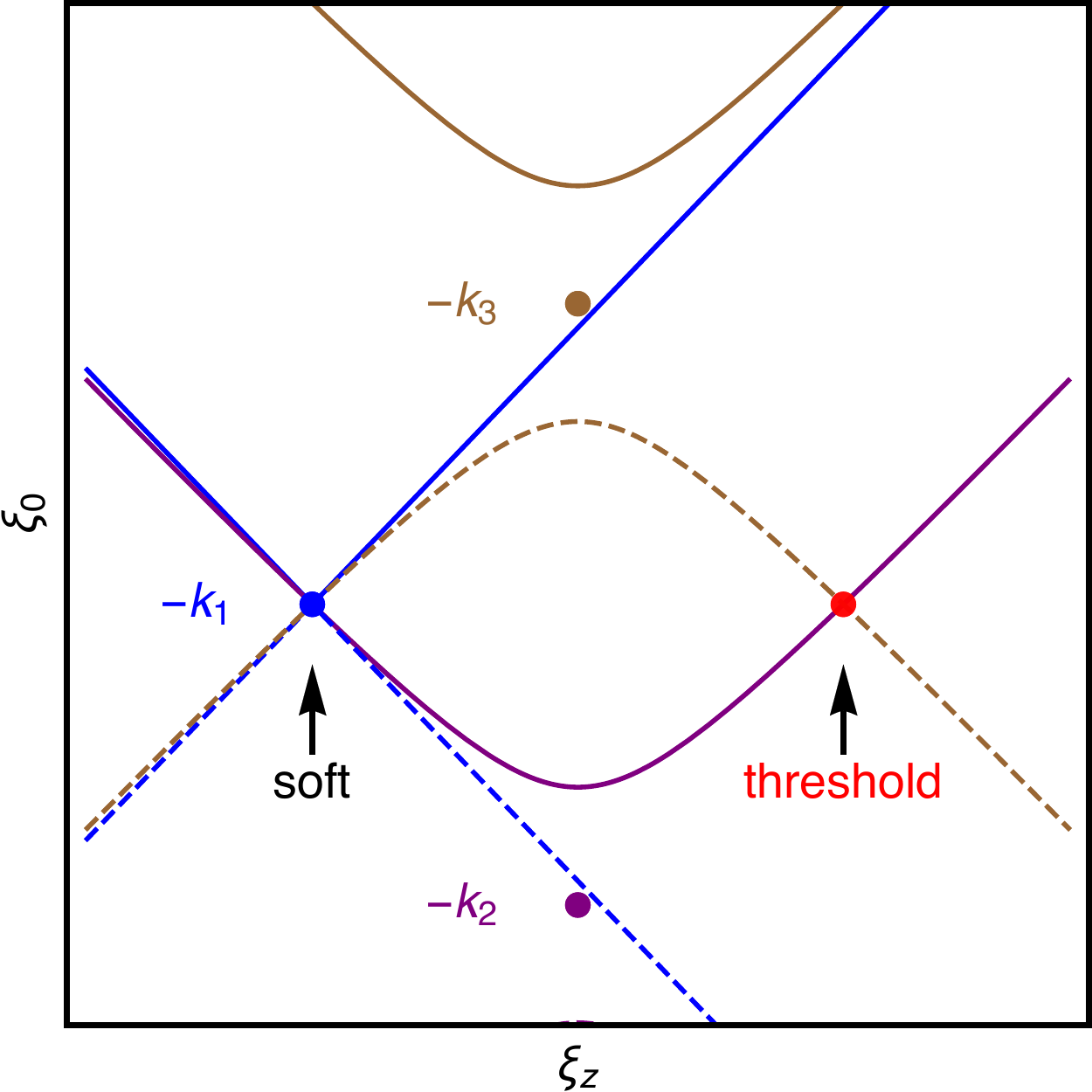} $\quad$
\includegraphics[width=0.45\textwidth]{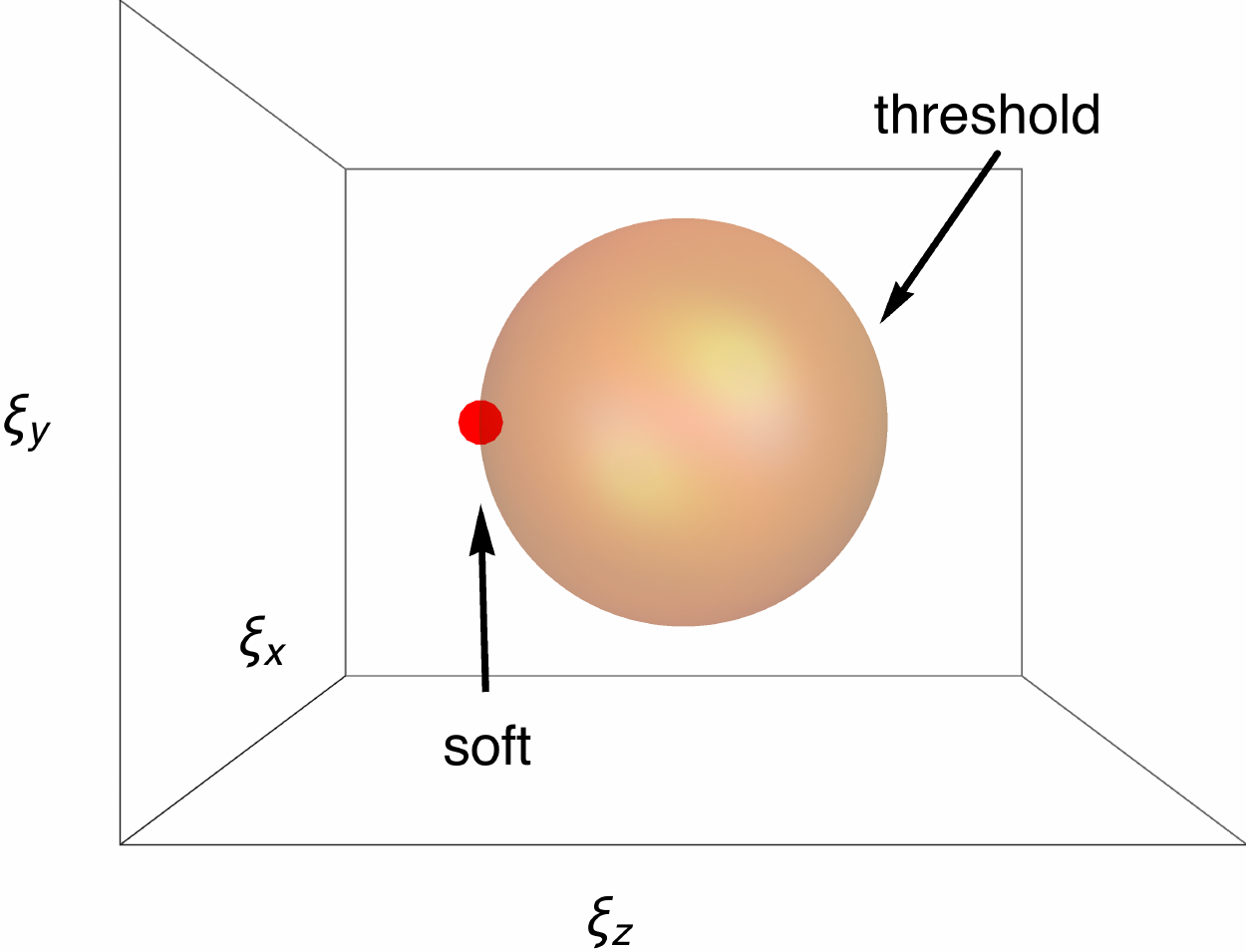}
\caption{On-shell hyperboloids of the massive three-point function in the loop coordinates $\ell^{\mu}= \sqrt{s_{12}}/2\, (\xi_0,\xi_x,\xi_y,\xi_z)$
in two dimensions (left plot); forward and backward on-shell hyperboloids are represented by solid and 
dashed lines, respectively. The intersection of on-shell hyperboloids leads to soft and threshold singularities
in the loop three-momentum space (right plot), collinear singularities are regulated by the mass. 
\label{fig:REGIONLightcone}}
\end{center}
\end{figure}

We present in this Section the first analytical application of LTD with massive particles.
In particular, we consider the scalar three-point function with one massless internal state and the remaining internal and outgoing particles with mass equal to $M$. The final-state on-shell momenta are labeled as $p_1$ and $p_2$, with $p_1^2=M^2=p_2^2$, and the incoming one is $p_3=p_1+p_2\equiv p_{12}$, by momentum conservation, with virtuality $p_3^2=s_{12}$. We consider $s_{12}>0$, i.e. we work in the time-like (TL) kinematical region. The internal momenta are $q_1 = \ell + p_1$,  $q_2 = \ell + p_{12}$ and $q_3 = \ell$, where $\ell$ is the loop momentum
(see Fig.~\ref{fig:NLOscalartoymodel}).
When the internal lines are set on-shell, the momentum $q_1$ is massless, whilst the other two are massive and fulfill
$q_2^2=M^2=q_3^2$. This is the master scalar configuration for the calculation of 
the QCD corrections to the physical case $A^* \to q \bar q (g)$ with massive quarks
that will be studied in Sec.~\ref{sec:physicalexamples}. We define
\beq
m = \frac{2 \, M}{\sqrt{s_{12}}}~, \qquad \beta = \sqrt{1-m^2},
\label{eq:mtdefinicion}
\eeq
as the normalised mass and velocity, respectively. The well-know result of this massive scalar three-point function 
is given by \cite{Ellis:2007qk,Rodrigo:1999qg}
\beqn
\nn L^{(1)}_{m > 0}(p_1,p_2,-p_3) &=& \int_\ell \, \prod_{i=1}^3 G_F(q_i) = -\frac{c_\Gamma}{s_{12}\, \beta} \, 
\bigg[ \ln{X_S} \bigg(-\frac{1}{\ep}-\frac{\ln{X_S}}{2} + 2\ln{1-X_S^2} \nn \\
&& + \ln{\frac{m^2}{4}}\bigg)+\li{X_S^2}+2\li{1-X_S} -\frac{\pi^2}{6} \bigg] + {\cal O}(\ep)~,
\label{eq:trianglemassive1}
\eeqn
with 
\beqn
X_{S} = - x_S - \imath 0 \, {\rm Sgn}(s_{12})~, \qquad x_S =\frac{1-\beta}{1+\beta}~,
\eeqn
and $c_{\Gamma}$ the one-loop volume factor
\beq
\cg = \frac{\Gamma(1+\ep) \, \Gamma^2(1-\ep)}{(4\pi)^{2-\ep} \, \Gamma(1-2\ep)}~.
\eeq 
Applying LTD, the dual representation of the scalar integral in~\Eq{eq:trianglemassive1} consists of three contributions
\beqn
L^{(1)}_{m > 0}(p_1,p_2,-p_3) &=& \sum_{i=1}^3 \, I_i \, ,
\label{eq:dualdecomposition}
\eeqn
with
\bea
I_1 &=& - \int_\ell \, \frac{\td{q_1}}{(2q_1\cdot p_2 - \imath 0)\,
(-2q_1\cdot p_1 + \imath 0)}~,
\label{CorteI1} \nn 
\\ I_2 &=& - \int_\ell \, \frac{\td{q_2}  }{(2 M^2-2q_2\cdot p_2 + \imath 0)\,  
(s_{12} -2q_2\cdot p_{12} + \imath 0)}~,
\label{CorteI2} \nn 
\\ I_3 &=& - \int_\ell \, \frac{\td{q_3}}{(2M^2+2q_3\cdot p_1 - \imath 0)\, 
(s_{12}+2q_3\cdot p_{12}- \imath 0)}~.
\label{CorteI3}
\eea
The corresponding on-shell hyperboloids are shown in Fig. \ref{fig:REGIONLightcone} (left). Due to the rotational symmetry, it is enough to  show the $(\ell_0,\ell_z)$ plane. As discussed in Ref. \cite{Buchta:2014dfa}, the intersection of on-shell hyperboloids is associated with multiple internal states becoming simultaneously on-shell. In this case, the forward on-shell hyperboloids of $G_F(q_1)$ and $G_F(q_2)$, and the backward one of $G_F(q_3)$ intersect in a single point: this leads to a soft singularity. The other intersection takes place between the forward on-shell hyperboloid of $G_F(q_2)$ with the backward of $G_F(q_3)$, which manifests as a threshold singularity in $I_2$. Notice that there are not collinear singularities, because the mass prevent that the on-shell hyperboloids degenerate into light-cones. In that situation, there would be \emph{extended} forward-backward intersections in the $(\ell_0,\ell_z)$ plane leading to collinear poles, as described in the massless case studied 
in Refs.~\cite{Hernandez-Pinto:2015ysa,Sborlini:2016gbr}.

Now, we shall explicitly compute these dual integrals. We choose first a proper reference frame to simplify the analytic expressions. Hence, we work in the centre-of-mass frame of $p_1$ and $p_2$, and parametrise the involved momenta as
\beq
p_1^{\mu} = \frac{\sqrt{s_{12}}}{2} \left(1, {\bf 0}_\perp, \beta \right)~, \quad \quad p_2^{\mu} = \frac{\sqrt{s_{12}}}{2} \left(1, {\bf 0}_\perp, -\beta \right)~.
\eeq
In order to describe the internal on-shell momenta, we must take into account that $q_1$ corresponds to a massless state ($q_1^2=0$), whilst $q_2$ and $q_3$ are associated with massive virtual particles ($q_2^2=M^2=q_3^2$). For this reason, we write
\beqn
q_1^{\mu} &=& \frac{\sqrt{s_{12}}}{2} \, \xi_{1,0} \,
\left(1, 2  \, \sqrt{v_1(1-v_1)} \, {\bf e}_{1,\perp}, 1-2v_1 \right) \, , \nn
\\ q_i^{\mu} &=& \frac{\sqrt{s_{12}}}{2} \, 
\left(\xi_{i,0}, 2 \, \xi_{i} \, \sqrt{v_i(1-v_i)} \, {\bf e}_{i,\perp}, \xi_{i} \, (1-2v_i) \right)~, 
\quad \xi_{i,0} = \sqrt{m^2+\xi_i^2}~, \quad i=\{2,3\} \ ,
\label{Trianguloqiparametrizacion}
\eeqn
where $\xi_{1,0}, \xi_2, \xi_3 \in [0,\infty)$ and $v_i \in [0,1]$ are the integration variables describing the modulus of the three-momentum and polar angle of the loop momenta, respectively. With these variables, the LTD transforms the loop integration measure into
\beq
\int_{\ell} \, \td{q_i} = s_{12} \, \int_0^{\infty} \frac{\xi_i^2}{\xi_{i,0}} \, d[\xi_i] \, \int_0^1 d[v_i]~,
\label{eq:loopmeasureMASS}
\eeq
with
\beq
d[\xi_i] = \frac{(4\pi)^{\ep-2}}{\Gamma(1-\ep)} \left(\frac{s_{12}}{\mu^2}\right)^{-\ep} \, \xi_i^{-2\ep} \, d\xi_i~, \qquad
d[v_i] = (v_i(1-v_i))^{-\ep} \, dv_i~,
\eeq
for the massive case ($i=2,3$), whilst it reduces to 
\beq
\int_{\ell} \, \td{q_1} = s_{12} \, \int_0^{\infty}  \xiu \,  d[\xiu] \, \int_0^1 d[v_1]~.
\label{eq:loopmeasureMASSLESS}
\eeq
for a massless state ($i=1$). Notice that $\xi_{1,0}=\xi_1$ since $q_1$ is massless. The integration of the loop momentum in the transverse plane, which is described by the unit vectors ${\bf e}_{i,\perp}$ is trivial. The scalar products of internal with external momenta are given by
\beqn
4 q_i\cdot p_1/s_{12} &=& \xi_{i,0}-\beta\,  \xi_{i} (1-2 v_i) ~,
\label{qip1} \nn
\\ 4 q_i\cdot p_2/s_{12} &=& \xi_{i,0}+\beta\, \xi_{i} (1-2 v_i)~,
\label{qip2}
\eeqn
which reduce to $2 q_1\cdot p_1/s_{12}=\xi_{1,0}v_1$ and $2 q_1\cdot p_2/s_{12}=\xi_{1,0}(1-v_1)$ for a massless on-shell state. Using these variables, the dual integrals in~\Eq{CorteI3} are rewritten as
\bea
I_1 &=& \frac{4}{s_{12}} \int \, \frac{\xi_{1,0}^{-1} \, d[\xi_{1,0}] \, d[v_1]}{1-(1-2 v_1)^2 \beta^2}~, \nn \\ 
I_2 &=& \frac{2}{s_{12}} \int  \frac{\xi_{2}^2 \, d[\xi_{2}] \, d[v_2]}{\xid \, 
\left(1-\xid+\imath 0 \right) \left(\xid+\beta \, \xi_{2} \, (1-2 v_2)-m^2\right)}~, \nn \\ 
I_3 &=&  -  \frac{2}{s_{12}} \int \frac{\xi_{3}^2 \, d[\xi_{3}] \, d[v_3]}{\xit \, \left(1+\xit \right) 
\left(\xit-\beta \, \xi_{3} (1-2 v_3)+m^2\right)}.
\label{eq:duals}
\eea
Notice that $I_{2}$ contains a threshold singularity at $\sqrt{m^2+\xi_2^2}=1$, i.e. $\xi_2=\beta \le 1$. These integrals can be calculated analytically 
to all orders in $\ep$ in the massless limit~\cite{Hernandez-Pinto:2015ysa}. In this limit, they read
\bea
I_1(m=0) &=& 0~, \nn
\\ I_2(m=0) &=& \cgt \, \frac{\mu^{2\ep}}{\ep^2} \, s_{12}^{-1-\ep} \, e^{\imath 2\pi \ep}~, \nn
\\ I_3(m=0) &=& \cgt \, \frac{\mu^{2\ep}}{\ep^2} \, s_{12}^{-1-\ep}~, 
\label{eq:duals0}
\eea
with $\cgt=\cg/ \cos(\pi \ep)$ the phase-space volume factor.
As expected, the sum of the three dual integrals in \Eq{eq:duals0} agrees with the well-known massless scalar three-point function.

The massive case is a bit more cumbersome because of the dependence on $m$. Again, $I_1$ vanishes because the energy integral factorises and it lacks of any characteristic scale. Actually, $I_1$ is singular both in the IR and UV. However, the sum of the three dual integrals and the 
equivalent original Feynman integral contain only soft divergences.
The other two dual integrals can be integrated in the angular variable analytically, thus keeping the exact $\ep$-dependence. This leads to
\beqn
I_2 &=& \frac{2 \Gamma^2(1-\ep)}{s_{12}\, \Gamma(2-2\ep)} \, \int d[\xi_2] \, \frac{ \xi_2^{2} \, 
_2F_1\left(1,1-\ep; 2-2 \ep; 2 \, \beta \, \xi_2 \, (\xid+\beta  \, \xi_2-m^2)^{-1}\right)}{\xid  \left(1-\xid+\imath 0\right)\left(\xid+\beta \, \xi_2-m^2\right)},
\nn \\ I_3 &=& -\frac{2 \Gamma^2(1-\ep)}{s_{12}\, \Gamma(2-2\ep)} \, \int d[\xi_3] \, \frac{\xi_3^{2} \, 
_2F_1\left(1,1-\ep; 2-2 \ep; -2 \, \beta \, \xi_3 \, (\xit - \beta  \, \xi_3+m^2)^{-1}\right)}{\xit \left(1+\xit\right) \left(\xit - \beta \, \xi_3+m^2\right)}~.
\label{eq:dualsinv}
\eeqn
However, an expansion in $\ep$ is necessary to integrate in the modulus of the loop three-momentum, which leads to a final result that includes corrections up to ${\cal O}(\ep^0)$. Besides that, the two dual integrals \Eq{eq:dualsinv} are singular only in the UV. Therefore, we introduce the following
expansion
\beq
I_i = \int \frac{d[\xi_i] }{\xi_{i,0}} \, g_i(\xi_i) =  \int \frac{d[\xi_i] }{\xi_{i,0}} \, g_\uv + 
 \int \left. \frac{d[\xi_i] }{\xi_{i,0}} \, \left( g_i(\xi_i) - g_\uv \right) \right|_{\ep=0}~,
\label{eq:intheUV}
\eeq
with $g_\uv = \lim_{\xi_i\to \infty} g_i (\xi_i)$. The first term in the r.h.s. of \Eq{eq:intheUV} gives the same result for both dual integrals, i.e.
\beq
\int \frac{d[\xi_i] }{\xi_{i,0}} \, g_\uv  = - 
\frac{\cg}{s_{12}} \frac{x_s^{-\ep} (1+x_s)^{1+2\ep} \, \Gamma(1-2\ep)}{2\ep (1-2\ep)  \Gamma^2(1-\ep)}
{}_2F_1(1,1-\ep; 2-2\ep, 1-x_s)~, \quad i=\{2,3\}~.
\eeq
For the second term, which is regular in the UV, we perform the change of variable
\beq
\xi_i = \frac{m}{2} \left(z - \frac{1}{z}\right)~,
\eeq
with $z \in [1, \infty)$. 
The total result for the real part of the two dual integrals, up to ${\cal O}(\ep)$, reads
\beqn
\nn {\rm Re}(I_2) &=&  \frac{\cg}{2 \beta\, s_{12}} \left(\frac{s_{12}}{\mu^2}\right)^{-\ep} \, 
\bigg[ \ln{x_S} \left( \frac{1}{\ep} - \frac{1}{2} \ln{x_S} - 2\ln{\beta} \right) 
- 2\li{x_S} - \frac{5\pi^2}{3} \bigg]+  {\cal O}(\ep)~, \label{I2dualmasivaanaliticaREAL}
\\ I_3 &=& \frac{\cg}{2 \beta\, s_{12}} \left(\frac{s_{12}}{\mu^2}\right)^{-\ep} \, 
\bigg[ \ln{x_S} \left( \frac{1}{\ep} - \frac{1}{2} \ln{x_S} - 2\ln{\beta} \right) 
- 2\li{x_S} + \frac{\pi^2}{3} \bigg]+  {\cal O}(\ep)~. \nn \\
\label{I3dualmasivaanalitica} 
\eeqn 
The dual integral $I_3$ is purely real, while the dual integral $I_2$ generates 
an imaginary component due to the intersection of the forward on-shell hyperboloid of $G_F(q_2)$ with 
the backward on-shell hyperboloid of $G_F(q_3)$. Its imaginary part can be calculated to all orders in $\ep$ from 
\bea
\imath \, {\rm Im}(I_2) &=& \frac{1}{2} \int_\ell G_D(q_2;q_1) \, \td{q_2} \, \td{-q_3}  \nn \\
&=& - \imath \pi \, \frac{2}{s_{12}} \int  \frac{ \delta\left(1-\xid \right) \, \xi_{2}^2 \, d[\xi_{2}] \, d[v_2]}{\xid \, 
 \left(\xid+\beta \, \xi_{2} \, (1-2 v_2)-m^2\right)}
 = \imath \, \frac{\cgt}{\beta \, s_{12}} \, \left(\frac{\beta^2\, s_{12}}{\mu^2}\right)^{-\ep}\frac{\sin(2\pi\ep)}{2 \ep^2}~,
\label{I2dualmasivaanaliticaIMAGINARIA}
\eeqn 
which is the expected result obtained through the application of the Cutkosky's rule. 
The sum of these contributions in \Eq{I3dualmasivaanalitica} and \Eq{I2dualmasivaanaliticaIMAGINARIA} agrees with the original 
Feynman integral in \Eq{eq:trianglemassive1}.

\section{Massive scalar decay rate in DREG}
\label{sec:scalarDREG}
\begin{figure}[t]
\begin{center}
\includegraphics[width=0.75\textwidth]{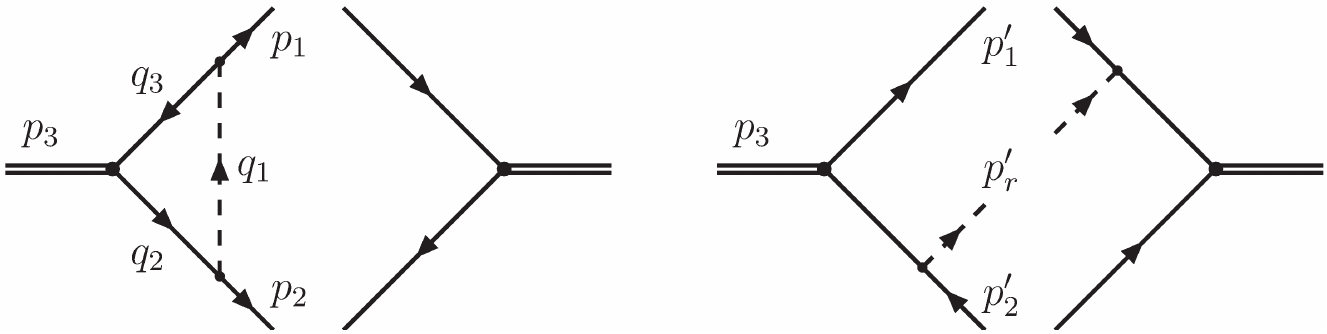}
\caption{Kinematic configuration of the NLO corrections to the decay process $\phi \to \phi+\phi$. The one-loop contribution is proportional to the scalar three-point function, with a virtual massless $\psi$ inside the loop (left). The real contribution is due to the interference terms originated by the emission of an on-shell massless particle $\psi$ (right). In this case, the momentum configuration is given by $p_3 \to p'_1+p'_2+p'_r$.
\label{fig:NLOscalartoymodel}}
\end{center}
\end{figure}

In order to establish a physical parallelism and understand the subtraction of IR singularities, we work in a simplified toy scalar model with a massive scalar particle $\phi$ which couples to a massless one, $\psi$. In concrete, we consider the decay process $\phi(p_3) \to \phi(p_1) + \phi(p_2)$, with $p_1^2=M^2=p_2^2$ (on-shell massive external particles) and $p_3^2=s_{12}$ (off-shell incoming particle). The Born-level decay rate is given by
\beqn
\Gamma^{(0)} &=& \frac{g^2}{2 \, \sqrt{s_{12}}} \, \int d\Phi_{1\to 2} \, ,
\eeqn
where $g$ is the coupling and $s_{12} > 4 M^2$ to guarantee the physical feasibility of the process. To compute the corresponding NLO correction, we need to add virtual (i.e. one-loop) and real (i.e. extra-radiation) contributions. We will assume the presence of only one massless particle inside the loop, as well as the emission of a massless real particle in the extra-radiation contribution.
The corresponding NLO diagrams are exhibited in Fig. \ref{fig:NLOscalartoymodel}~\footnote{This decay rate does not correspond to any physical theory, since the full set of Feynman diagrams has not be taken into account. However, for illustrative purposes, it is enough to restrict the following discussion to virtual and real contributions with a similar topology~\cite{Hernandez-Pinto:2015ysa,Sborlini:2016gbr}.}. 

Let's start with the virtual part, which we assume proportional to the scalar three-point function, i.e.
\beqn 
\Gamma_{\v}^{(1)}  &=& \frac{1}{2\, \sqrt{s_{12}}} \, \int d\Phi_{1\to 2} \, 2 \, \re \, \la \M{0} | \M{1} \ra= - \Gamma^{(0)} \,  2\, g^2 \, s_{12}\, \re \, L^{(1)}_{m > 0}(p_1,p_2,-p_3) ~.
\label{eq:sigmavscalar}
\eeqn
Since $s_{12}>0$, the virtual decay rate is given from \Eq{I3dualmasivaanalitica} as a function of $x_S$
\beqn
\nn \Gamma_{\v}^{(1)} &=&  \Gamma^{(0)} \, \frac{\cg \, 4 \, g^2}{\beta} \left(\frac{s_{12}}{\mu^2}\right)^{-\ep} \, 
\bigg[ \ln{x_S} \left( -\frac{1}{2\ep} + \frac{1}{4} \ln{x_S} + \ln{\beta} \right) 
+ \li{x_S} + \frac{\pi^2}{3} \bigg]+  {\cal O}(\ep)~. \nn \\
\label{eq:sigmavscalar2} 
\eeqn 
In order to calculate the totally inclusive decay rate, we have to consider the real emission process. In this particular toy-example the cancellation of IR-singularities is achieved by including only the interference terms originated in the process $\phi(p_3) \to \phi(p_1')+\phi(p_2')+\psi(p_r')$, as depicted schematically in Fig. \ref{fig:NLOscalartoymodel}. Explicitly, 
\beqn
\nn \Gamma^{(1)}_{\r} &=& \frac{1}{2 \, \sqrt{s_{12}}} \, \int d\Phi_{1\to 3} \, 2 \, \re \, \la \M{0}_{2r} | \M{0}_{1r} \ra = \frac{g^4}{2\, \sqrt{s_{12}}} \int  d\Phi_{1\to 3} \frac{2\ s_{12}}{(2\, p_1' \cdot p_r')(2\, p_2' \cdot p_r')} \, 
\\ &=&  \Gamma^{(0)} \, 2  g^2 \,  \frac{(4\pi)^{\ep-2}}{\Gamma(1-\ep)} \, 
\left( \frac{s_{12}}{\mu^2}\right)^{-\ep} \,  
\beta^{-1+2\ep} \, \int \theta(h_p) \, h_p^{-\ep}\, \frac{dy_{1r}'  \, dy_{2r}'}{y_{1r}' \, y_{2r}'} \, ,
\label{eq:GammaRscalar}
\eeqn
where we used the definition of the massive three-body phase space in \Eq{eq:PHASESPACE} and $y_{ir}'=2 \, p'_i \cdot p'_r /s_{12}$. To compute this integral, we apply the change of variables suggested in Appendix \ref{app:PSformulae}, which allows to factorise the energy and the angular dependence of the integrand. By using \Eq{eq:gzdef}, we obtain
\beqn
\nn \Gamma^{(1)}_{\r} &=& \Gamma^{(0)} \, 2  g^2 \,  \frac{(4\pi)^{\ep-2}}{\Gamma(1-\ep)} \, 
\left( \frac{s_{12}}{\mu^2}\right)^{-\ep} \,  \beta^{-1+2\ep}\, (1+x_S)^{6\ep} 
\\ &\times& \int_{x_S}^{x_S^{-1}} dz \, \frac{z^{-1+2\ep}(1+z)^{2\ep}}{(z-x_S)^{3\ep} (1-x_S \, z)^{3\ep}} \, \int_0^{1}dw \, w^{-1-2\ep} (1-w)^{-\ep}.
\eeqn
The integration in $w$ can be trivially performed, and it leads to the appearance of an $\epsilon$-pole. The integral in $z$ is finite if $x_S>0$ (i.e. in the massive case), thus we expand the integrand in $\ep$ before the integration. The resulting expression is
\beqn
\nn \Gamma^{(1)}_{\r}&=&  \Gamma^{(0)} \, \frac{ \cg \, 4  \, g^2}{\beta}  \,
\left( \frac{s_{12}}{\mu^2}\right)^{-\ep} \, \bigg[ \ln{x_S} \left( \frac{1}{2\ep} - \frac{1}{4} \ln{x_S} + \ln{1+x_S} + \ln{1-x_S^2} \right) \nn \\
&+& \li{x_S} + \li{x_S^2} - \frac{\pi ^2}{3} \bigg]+ \, {\cal O}(\ep)~. 
\label{eq:decayscalarR}
\eeqn
Putting together the virtual and real contributions from \Eq{eq:sigmavscalar2} and \Eq{eq:decayscalarR}, we get
\beqn
\Gamma^{(1)} &=&   \Gamma^{(0)} \, \frac{4 \, a}{\beta} \,  \left[ \ln{x_S}\left(\ln{1-x_S}+\ln{1-x_S^2}\right) + 2 \li{x_S} + \li{x_S^2} \right]
+ {\cal O}(\ep)~,
\label{eq:TOYresultadoDREG}
\eeqn
with $a=g^2/(4\pi)^2$. The purpose of the following discussion will be the derivation of a purely four-dimensional representation of this result, through the local cancellation of all the IR divergences present in both real and virtual contributions. It is crucial to emphasise that this cancellation of IR singularities at integrand level is achieved by a suitable mapping of momenta.

\section{Phase-space partition and real-virtual mapping with massive particles}
\label{sec:realvirtualmap}
The first ingredient that we need to introduce is a complete partition of the real 
phase-space~\cite{Hernandez-Pinto:2015ysa,Sborlini:2016gbr}, in such a way that each 
individual region of that partition contains a single soft, collinear or quasi-collinear configuration.  
The quasi-collinear configurations are those in which a massless particle becomes collinear with a massive one~\cite{Catani:2002hc}. 
In that case the mass acts as an IR-regulator and the collinear $\ep$-poles that appear in DREG in the massless case 
are transformed into finite logarithmic terms in the mass. These logarithmic contributions are cancelled in the total cross-section but 
the massless and the $\ep \to 0$ limits do not commute for the virtual and real corrections separately.  
Thus, we split the real phase-space by defining the domains
\beqn
{\cal R}_{i} &=& \{y'_{ir} < {\rm min}(y'_{jk}) \}~, \qquad  \sum {\cal R}_i = 1~.
\label{eq:regiones}
\eeqn
For instance, ${\cal R}_i$ selects configurations with $p_i' \parallel p_r'$ or close to collinear, excluding the remaining ones. 
In particular, this partition reduces to 
\beq
\theta(y_{2r}'-y_{1r}') + \theta(y_{1r}'-y_{2r}')=1~,
\label{eq:simplestsplit}
\eeq
for the simplest $1\to 3$ scenario. 
The definition of the phase-space partition in \Eq{eq:regiones} is the same that we would use in the massless case~\cite{Sborlini:2016gbr}; 
the only difference is that now some of the external momenta are massive. That partition, together with the well-motivated mapping of momenta 
that will be presented in the following, ensures a smooth massless limit, therefore an integrand level cancellation 
of the logarithmic dependences in the mass arising from the quasi-collinear configurations of the virtual and real corrections, and thus a more stable 
numerical implementation of the method. 

Then, we shall define a proper momentum mapping in each region to match the 
singular behaviour of the real and the dual integrands.
In order to properly combine real and virtual contributions at integrand level, we need to generate the $N+1$ on-shell kinematics by making use of the $N$-parton Born-level process and the on-shell loop momenta. One of the main difficulties in constructing a momentum mapping with massive particles is that the on-shell conditions lead to quadratic equations in the mapping parameters. However, it is very well-known that massive vectors can be expressed in terms of two massless momenta. Then, we can exploit this property to simplify the mapping equations. For the case 
of a pair of particles of the same mass, the corresponding massive momenta can be written as 
\beq
p_1^{\mu} = \beta_{+}\hat{p}_1^{\mu}+\beta_{-}\hat{p}_2^{\mu}~, \quad \quad p_2^{\mu} = \beta_{-}\hat{p}_1^{\mu}+\beta_{+}\hat{p}_2^{\mu}~,
\label{eq:momentap1andp2}
\eeq
with $\hat p_1^2=\hat p_2^2 = 0$ and $\beta_{\pm}=(1\pm \beta)/2$. 
Moreover, the massless momenta fulfil the following useful identities 
\beq
2 \, \hat p_1\cdot \hat p_2 = s_{12}~, \qquad \hat p_1^\mu +\hat p_2^\mu = p_1^\mu+p_2^\mu~.
\eeq
In their centre-of-mass frame, these massless momenta are simply given by 
\beq
\hat{p}_1^{\mu} = \frac{\sqrt{s_{12}}}{2}\left(1, {\bf 0}_\perp, 1\right) \, , \quad \quad \hat{p}_2^{\mu} = \frac{\sqrt{s_{12}}}{2}\left(1, {\bf 0}_\perp, -1\right)~.
\label{eq:momentap1andp2hat}
\eeq

Going back to the toy example in Sec.~\ref{sec:scalarDREG},
let's start with the first region where ${\cal R}_1=1$ (i.e. $y_{1r}' < y_{2r}'$).
Motivated by the factorisation properties of QCD in the quasi-collinear limit and the momentum decomposition 
in \Eq{eq:momentap1andp2}, we propose the following mapping with $q_1^2=0$
\bea
\nn && p_r'^\mu = q_1^\mu~,
\\ \nn && p_1'^\mu = (1-\alpha_1) \, \hat p_1^\mu + (1-\gamma_1) \, \hat p_2^\mu - q_1^\mu~, 
\\ && p_2'^\mu = \alpha_1 \, \hat p_1^\mu + \gamma_1 \, \hat p_2^\mu~,
\label{eq:mapping1}
\eea
which fulfils the momentum conservation constraint by construction. 
As in the massless case, the momentum $p_2'$ acts as the spectator of the splitting process 
and is used to balance momentum conservation. The emitters have momenta $p_1$ and $p_1'$, and have the same mass. 
Although restricted to three final-state particles, the momentum mapping in \Eq{eq:mapping1} 
can easily be generalised to the multipartonic case, with $p_k' = p_k$ for $k\ne 1,2,r$. 
The parameters $\alpha_1$ and $\gamma_1$ are determined from the two on-shell conditions 
\bea
&& (p_1')^2 = (1-\alpha_1) (1-\gamma_1) \, s_{12} - 2 q_1\cdot((1-\alpha_1)  \, \hat p_1+(1-\gamma_1) \, \hat p_2) = M^2~, \nn \\
&& (p_2')^2 = \alpha_1 \, \gamma_1 \, s_{12} = M^2,
\label{eq:onshellconditions1}
\eea
whose explicit solutions are
\beqn
\nn \alpha_1 &=& \frac{1-\xiu-\sqrt{ (1-\xiu)^2-m^2(1-\xi_{1,0}+v_1(1-v_1)\, \xiu^2)}}{2\,(1-v_1\, \xiu)}~, 
\\ \gamma_1 &=& \frac{1-\xiu+\sqrt{ (1-\xiu)^2-m^2(1-\xi_{1,0}+v_1(1-v_1)\, \xiu^2)}}{2\,(1-(1-v_1)\, \xiu)}~,
\label{eq:MAP1soluciones}
\eeqn
whilst $(p_r')^2=0$ by construction, since $q_1^2=0$. Due to the fact that we are dealing with quadratic equations, there are two sets of solutions. 
The solution in \Eq{eq:MAP1soluciones} is compatible with the soft limit; it recovers the Born-level kinematics when $\xi_{1,0} \to 0$. 
In that limit, $(\alpha_1, \gamma_1) \to (\beta_-, \beta_+)$ and therefore $(p_1',p_2') \to (p_1,p_2)$. 
Also, it properly reduces to the massless parametrisation defined in Refs. \cite{Hernandez-Pinto:2015ysa,Sborlini:2016gbr}, i.e. 
if $m \to 0$, we have
\beqn
\alpha_1 \to 0~, \qquad \gamma_1 \to \frac{1-\xi_{1,0}}{1-(1-v_1)\xi_{1,0}}~.
\label{eq:MAP1solucionesMASSLESS}
\eeqn
Using these definitions, the kinematical invariants $y'_{ij}$ become
\beqn
y'_{1r} &=& \frac{\xiu}{1-(1-v_1)\,\xiu} \left( v_1 + \alpha_1 \, (1-2v_1) \right)~, \nn \\ 
y'_{2r} &=&\frac{\xiu}{1-(1-v_1)\,\xiu} \left( (1-v_1)(1-\xiu) - \alpha_1 \, (1-2v_1) \right)~, \nn \\ 
y'_{12} &=& 1-\xiu-\frac{m^2}{2}~,
\eeqn
which fullfil $y'_{12}+y'_{1r}+y'_{2r}=1-m^2/2$.  Again, we recover easily the massless expressions \cite{Hernandez-Pinto:2015ysa,Sborlini:2016gbr} with $\alpha_1 = 0$. In order to improve the presentation of the results, it is also convenient to express the mass in terms of $\alpha_1$,
\beq
m^2 = \frac{4\alpha_1 \, (1-\xiu-\alpha_1 \, (1-v_1 \, \xiu))}{1-(1-v_1)\, \xiu}~.
\label{eq:masaALPHA1}
\eeq
Then, we compute the associated Jacobian in the physically allowed region (i.e. those points belonging to the domain ${\cal R}_1$), which is given by
\beq
{\cal J}_1(\xiu,v_1) = \frac{\xiu \, (1-\xiu-\alpha_1 \, (2-\xiu))^2}{(1-(1-v_1) \, \xiu)^2(1-\xiu-2\alpha_1 \, (1-v_1\, \xiu))}~,
\label{eq:Jacobiano1}
\eeq
with $dy'_{1r} \, dy' _{2r} = {\cal J}_1(\xiu,v_1) \, d\xiu \, dv_1$.
Notice that this expression is apparently free of square roots, since the mass dependence was rewritten in 
terms of $\alpha_1$, as suggested in \Eq{eq:masaALPHA1}. 

\begin{figure}[t]
\begin{center}
\includegraphics[width=7cm]{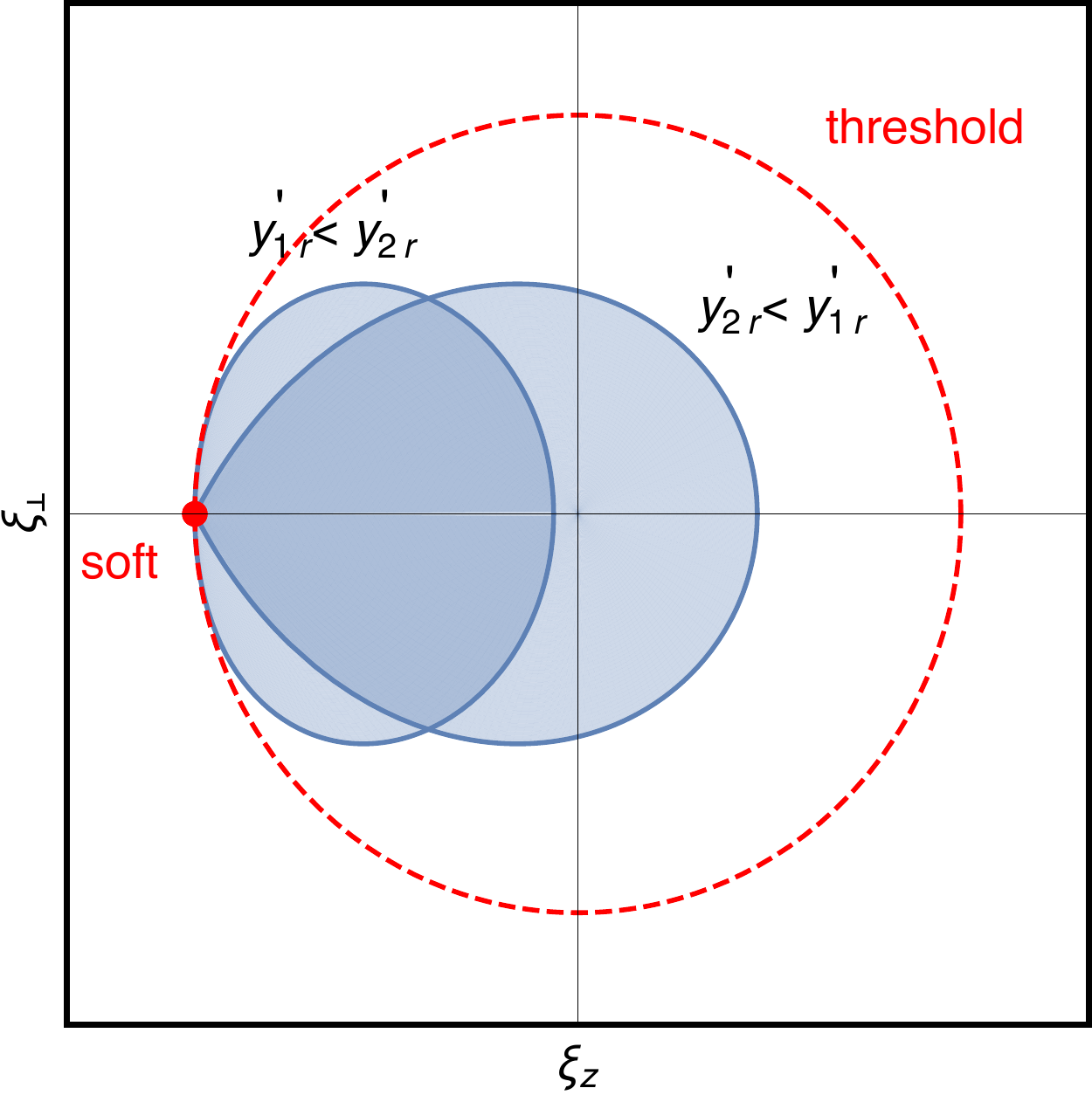}
\caption{\label{fig:dualregions} The dual integration regions in the loop three-momentum space, with $\xi_\perp = \sqrt{\xi_x^2+\xi_y^2}$.}
\end{center}
\end{figure}

On the other hand, we have to express ${\cal R}_1$ in terms of the dual variables. If we use the mapping given in \Eq{eq:mapping1}, we obtain
\beqn
{\cal R}_1(\xiu, v_1) &=& \theta(1-2v_1) \, \theta \left( \frac{1-2v_1}{1-v_1} \left(1-\frac{1-\sqrt{1-4m^2\, v_1 (1-v_1)}}{2v_1}\right) - \xiu \right)~,
\label{eq:R1xiv}
\eeqn
which is the characteristic function associated to the domain ${\cal R}_1$. In Fig.~\ref{fig:dualregions}, we show this domain 
in the loop three-momentum space. The massless limit agrees with the expected result. 
Moreover, the three-body phase-space limits defined by the condition $h_p=0$ are simply determined by $v_1=0$.

In the complementary region, where ${\cal R}_2=1$ (i.e. $y_{2r}' < y_{1r}'$) with $q_2^2 = M^2$, 
the mapping is defined by~\footnote{We have exchanged the role of the radiated particle and the emitter
to keep $p_2'$ massive.}
\bea
\nn && p_r'^\mu =(1-\gamma_2) \, \hat p_1^\mu + (1-\alpha_2) \, \hat p_2^\mu - q_2^\mu~,
\\ \nn && p_1'^\mu = \gamma_2 \, \hat p_1^\mu + \alpha_2 \, \hat p_2^\mu~, 
\\ && p_2'^\mu = q_2^\mu~,
\label{eq:mapping2}
\eea
where
\bea
&& (p_1')^2 = \alpha_2\, \gamma_2 \,  s_{12} = M^2~, \nn \\
&& (p_r')^2 = M^2 + (1-\alpha_2) \, (1- \gamma_2) \, s_{12} - 2 q_2 \cdot((1-\gamma_2) \, \hat p_1+(1-\alpha_2) \, \hat p_2) = 0,
\label{eq:onshellconditions2}
\eea
are the associated on-shell conditions. In this case, the condition $(p_2')^2=M^2$ is fulfilled by construction. Solving the system and selecting the physical solution, we get
\beqn
\nn \alpha_2 &=& \frac{1-\xid+m^2/2- \sqrt{(1-\xi_{2,0})^2-m^2\, v_2(1-v_2)\, \xi_2^2}}{2-(1-2 v_2)\, \xi_2-\xid} \, ,
\\ \gamma_2 &=& \frac{1-\xid+m^2/2+\sqrt{(1-\xi_{2,0})^2-m^2\, v_2(1-v_2)\, \xi_2^2}}{2+(1-2 v_2)\, \xi_2-\xid}\, .
\label{eq:MAP2soluciones} 
\eeqn
In order to check the consistency of this solution, we consider the massless limit, obtaining
\beqn
\alpha_2 \to 0~, \qquad \gamma_2 \to \frac{1-\xi_2}{1-v_2 \, \xi_2},
\label{eq:MAP2solucionesMASSLESS}
\eeqn
which implies that the parametrisation reduces to the expected one. 
On the other hand, the two-body invariants are given by
\beqn
y'_{1r} &=& 1 - \xid~,  \nn \\ 
y'_{2r} &=& \frac{\xid+(1-2\, v_2)\, (1-2\, \alpha_2)\, \xi_2 - m^2}{2+(1-2\, v_2)\, \xi_2-\xid}~, \nn \\
y'_{12} &=& \xi_{2,0} - \frac{m^2}{2} - y_{2r}'~, 
 \eeqn
and, since this mapping will be used in the region of the real phase-space defined by ${\cal R}_2$, 
we rewrite the associated characteristic function as 
\beqn
{\cal R}_2(\xi_{2},v_2) &=& \theta \left( \beta^2\, \left(\left(1+\sqrt{(1-v_2)\, (1-m^2 \, v_2)}\right)^2-m^2 \, v_2^2\right)^{-1/2}-\xi_2 \right)~.
\label{eq:R2xiv}
\eeqn 
The corresponding domain in the loop three-momentum space is also shown in Fig.~\ref{fig:dualregions}.
The associated Jacobian is given by
\beqn
{\cal J}_2(\xi_{2},v_2) &=& \frac{4\xi_2^2 \,  \left( 1-\xid +m^2/2 - \alpha_2 \, (2-\xid)\right)^2}
{\xi_{2,0}\, (2+(1-2v_2)\, \xi_2-\xid)^2 \, \left( 1-\xid +m^2/2- \alpha_2 \, (2-(1-2v_2)\, \xi_2-\xid) \right)}~, \nn \\
\label{eq:Jacobiano2}
\eeqn
with $dy'_{1r} \, dy' _{2r} = {\cal J}_2(\xi_2,v_2) \, d\xi_2 \, dv_2$, and we made use of the identity
\beqn
m^2 &=& \frac{4\alpha_2 \left(2 (1-\xid) - \alpha_2 (2-(1-2v_2) \, \xi_2-\xid)\right)}{2+(1-2v_2)\, \xi_2-\xid-4\alpha_2} \, ,
\eeqn
to simplify the expressions.

\subsection{General momentum mapping}
\label{app:generalmapping}
The momentum mappings previously presented can easily be extended to the most general multipartonic case in which the emitter and the spectator have different masses: $p_i^2=m_i^2$ and $p_j^2=m_j^2$, respectively. The decomposition of their momenta in terms of two massless momenta ($\hat p_i^2 = \hat p_j^2=0$) is given by 
\bea
&& p_i^\mu = \beta_+ \, \hat p_i^\mu +  \beta_- \, \hat p_j^\mu~, \nn \\ 
&& p_j^\mu = (1-\beta_+) \, \hat p_i^\mu +  (1-\beta_-) \, \hat p_j^\mu~,
\eea
with
\beq
\beta_\pm = \frac{s_{ij}+m_i^2-m_j^2 \pm \lambda(s_{ij},m_i^2,m_j^2)}{2 s_{ij}}~,
\eeq
where $\lambda(s_{ij},m_i^2,m_j^2)=\sqrt{(s_{ij}-(m_i-m_j)^2) (s_{ij}-(m_i+m_j)^2)}$ is the usual Kall\'en function.
The massless momenta fulfil the useful condition $\hat p_i + \hat p_j = p_i+p_j$.
The mapping with the momenta of the real process is formally equal to the mapping already considered in \Eq{eq:mapping1}, i.e.
\bea
&& p_r'^\mu = q_i^\mu  \nn \\
&& p_i'^\mu = (1-\alpha_i) \, \hat p_i^\mu + (1-\gamma_i) \, \hat p_j^\mu-q_i^\mu~, \nn \\
&& p_j'^\mu = \alpha_i \, \hat p_i^\mu + \gamma_i \, \hat p_j^\mu~, \nn \\
&& p_k'^\mu = p_k^\mu~, \qquad k\ne i,j,r~.
\label{eq:generalmapping}
\eea
It leads to the on-shell conditions
\bea
&& (p_i')^2 =  (1-\alpha_i) (1- \gamma_i)\, s_{ij} - 2q_i \cdot \left( (1-\alpha_i) \, \hat p_i + (1-\gamma_i) \, \hat p_j \right) + m_r^2= (m_i')^2~,\nn \\
&& (p_j')^2 = \alpha_i \, \gamma_i \, s_{ij} = m_j^2~. 
\label{eq:generalonshell}
\eea
In \Eq{eq:generalonshell}, we have imposed that the spectator and the radiated particle have the same 
flavour (and, thus, the same mass) in the virtual and 
real processes; $p_j^2=(p_j' )^2=m_j^2$ and $q_i^2=(p_r')^2=m_r^2$, respectively. The emitter, however, might change flavour,
$(p_i')^2 = (m_i')^2 \ne m_i^2$. This situation occurs, for instance, when a gluon splits into a massive quark-antiquark pair. 
The solution to \Eq{eq:generalonshell} for the parameters of the mapping reads
\bea
&& \alpha_i = \frac{(p_{ij}-q_i)^2+m_j^2-(m_i')^2- \Lambda_{ij}}{2(s_{ij}-2 q_i\cdot \hat p_i)}~, \nn \\
&& \gamma_i = \frac{(p_{ij}-q_i)^2+m_j^2-(m_i')^2+ \Lambda_{ij}}{2(s_{ij}-2 q_i\cdot \hat p_j)}~,
\eea
with 
\beq
\Lambda_{ij} = \sqrt{ ( (p_{ij}-q_i)^2+m_j^2-(m_i')^2)^2 - \frac{4m_j^2}{s_{ij}} (s_{ij}-2q_i\cdot \hat p_i) (s_{ij}-2q_i\cdot \hat p_j)}~.
\eeq
The momentum mapping in \Eq{eq:generalmapping} has an smooth limit whenever any of the involved particles 
becomes massless; in particular, if the spectator is a massless particle, then $\alpha_i = 0$.

\section{Massive scalar decay rate from four-dimensional unsubstraction}
\label{sec:unsubtraction}
In this section we illustrate the method of LTD four-dimensional unsubstraction~\cite{Hernandez-Pinto:2015ysa,Sborlini:2016gbr} with the massive toy example presented in Sec.~\ref{sec:scalarDREG}. All the necessary ingredients have been presented in the previous sections.
We combine at integrand level the dual loop contributions (Sec.~\ref{sec:massivetriangle}) with the real-radiation terms (Sec.~\ref{sec:scalarDREG})
with the help of the momentum mappings defined in Sec.~\ref{sec:realvirtualmap}. Since the sum of all the contributions 
is UV and IR finite, the final result is free of $\ep$-poles. We would like to emphasise that, in a generic situation, this assertion is not enough to guarantee the integrability of the expressions in four-dimensions. However, by virtue of the momentum mappings and the unification of the dual coordinates, LTD leads naturally to a local cancellation of divergences and the limit $\ep \to 0$ can be considered at \emph{integrand} level. 

The LTD representation of the virtual decay rate in this toy example is given by 
\beqn
\G{1}_{\v} &=& \frac{1}{2\, \sqrt{s_{12}}} \, \sum_{i=1}^3
\int \, d\Phi_{1\to 2} \, 2 \, \re \, \la \M{0} | \M{1} (\td{q_i}) \ra~, 
\label{eq:toyV}
\eeqn
with
\beqn
\la \M{0} | \M{1}(\td{q_i}) \ra &=& -g^4 \, s_{12} \, I_i~, 
\eeqn
where the dual integrals $I_i$ are defined in~\Eq{eq:duals}, and we take the integration measure exactly with $\ep=0$.
In order to ensure the cross-cancellation of spurious singularities and get a direct $\ep=0$ limit we must
rewrite all the on-shell momenta in terms of the same coordinate system. This change of variables is explained in 
Appendix~\ref{app:unificationcoordinates}. With that change of variables the virtual decay rate 
in \Eq{eq:toyV} becomes a single unconstrained integral in the loop three-momentum. 

We also consider the real contribution given by \Eq{eq:GammaRscalar}. First, we split the real three-body 
phase-space according to \Eq{eq:simplestsplit}, and define 
\beq
\widetilde{\Gamma}^{(1)}_{\r,i} = \frac{1}{2 \sqrt{s_{12}}} \, \int\, d\Phi_{1\to 3} \, 2 \, \re \la \M{0}_{2r} | \M{0}_{1r} \ra \, 
{\cal R}_i\left(y'_{ir} < y'_{jr}\right) \, , \qquad i,j=\{1,2\}~,
\eeq 
that obviously fulfill
\beq
\Gamma^{(1)}_{\r} = \widetilde{\Gamma}^{(1)}_{\r,1}+ \widetilde{\Gamma}^{(1)}_{\r,2}~.
\label{eq:toyR}
\eeq
Second, in each region of the real phase-space we apply one of the momentum mappings defined in 
\Eq{eq:mapping1} and \Eq{eq:mapping2}, respectively. The main advantage of these mappings is that they 
are optimised to deal smoothly with the massless limit in each of the two regions. 
Thus, we rewrite the real contributions in terms of the loop variables and we obtain
\beqn
\widetilde{\Gamma}_{\r,1}^{(1)} &=& \G{0} \, \frac{2 a}{\beta} \, \int \, d\xiu \, dv_1 \, 
\frac{ {\cal R}_1(\xiu, v_1) \,  {\cal J}_1(\xiu, v_1) \,  (1-\xi_{1,0} (1-v_1))^2}{\xiu^2 \, (v_1+\alpha_1(1-2 v_1)) ((1-v_1)(1-\xiu)-\alpha_1(1-2v_1))}~,
\label{eq:GammaTilde1TOY} \\ 
\widetilde{\Gamma}_{\r,2}^{(1)} &=& \G{0} \, \frac{2 a}{\beta} \, \int \, d\xi_{2} \, dv_2  \, 
\frac{ {\cal R}_2(\xi_2,v_2)\, {\cal J}_2(\xi_2,v_2) \, (2+(1-2v_2) \, \xi_2 - \xid)}{(1-\xid) (\xid + (1-2 v_2)\, (1-2 \alpha_2) \, \xi_2 - m^2)}\, ,
\label{eq:GammaTilde2TOY}
\eeqn
where the Jacobians of the respective transformations are given by \Eq{eq:Jacobiano1} and \Eq{eq:Jacobiano2}, 
and the integration domains, which are restricted by the functions ${\cal R}_i$ from \Eq{eq:R1xiv} and \Eq{eq:R2xiv},
are shown in Fig.~\ref{fig:dualregions}.
Again, after applying  the change of variables defined in Appendix~\ref{app:unificationcoordinates} to bring the two real contributions 
to a common coordinate system, we can consistently take the limit $\ep=0$, directly at integrand level.

The sum of all the virtual and real contributions, Eqs.~(\ref{eq:toyV}),  (\ref{eq:GammaTilde1TOY}) and (\ref{eq:GammaTilde2TOY}), 
is a finite function in the $\ep=0$ limit because all the IR singularities are cancelled locally in 
the loop three-momentum space at integrand level. The virtual contribution, however, contains a threshold 
singularity at $\xi_2 = \beta$. This singularity is integrable and can be treated numerically 
by contour deformation~\cite{Buchta:2015xda,Buchta:2015wna,Buchta:2015jea}.
For the toy scalar model and the physical examples that we are considering in this article there is a simplest solution:
we can compactify the high-energy region with $\xi > \beta$ into the unit sphere by using a change of variables. Explicitly, 
since the integrand is a function of the modulus of the three-momentum and the polar angle, then 
\beq
\int_0^{\infty} d\xi \, g(\xi,v) = \beta \, \int_0^1 dx \left[ g(\beta \, x, v) + x^{-2}\, g(\beta \, x^{-1},v) \right]~,
\label{eq:inverse}
\eeq
where the threshold singularity has been mapped into to the upper end-point, namely $x=1$.
This approach is very efficient for the numerical implementation. 

\begin{figure}[t]
\begin{center}
\includegraphics[width=0.5\textwidth]{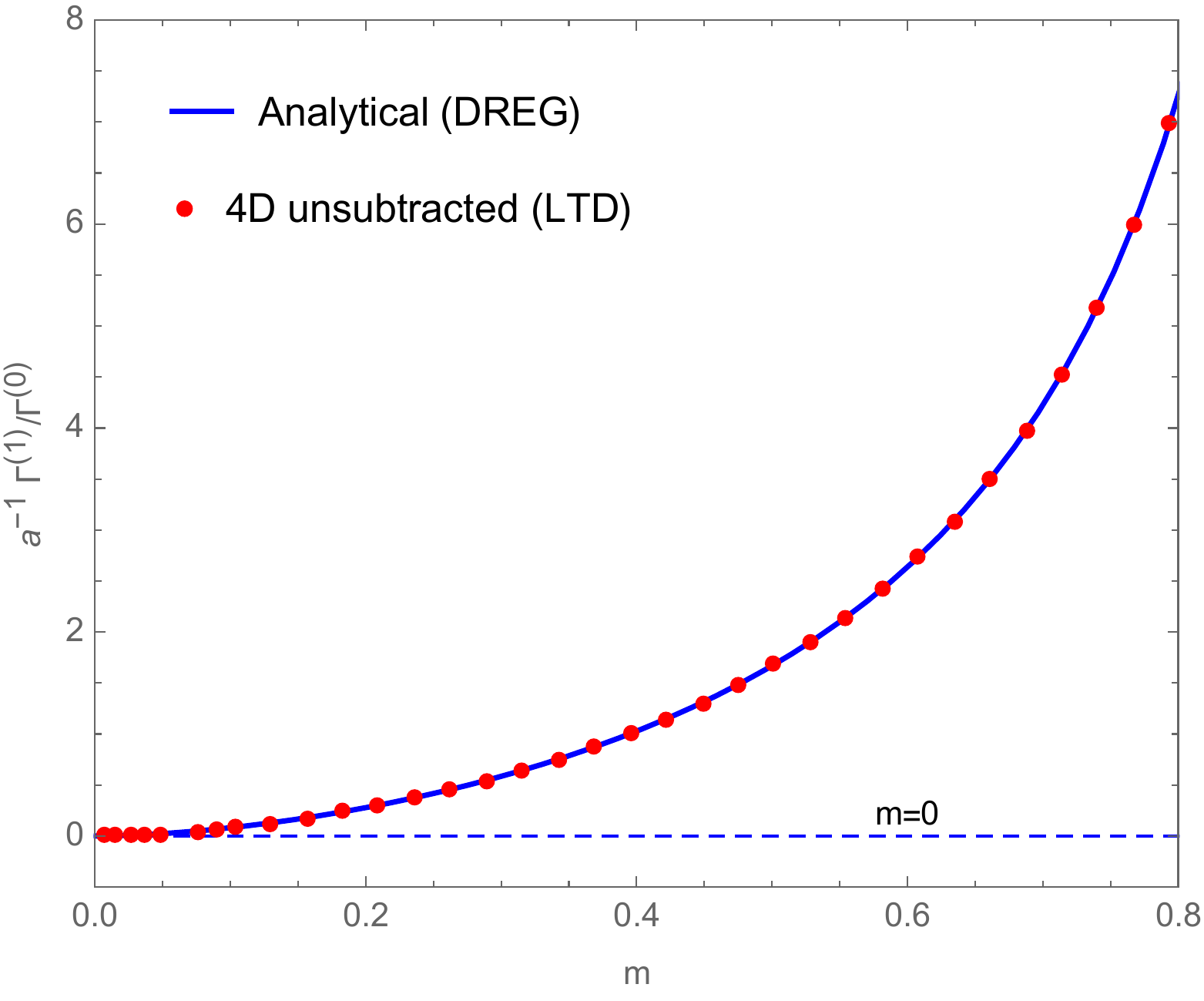}
\caption{Total decay rate at NLO normalised to the leading order for the toy scalar example,  $a^{-1} \Gamma^{(1)}/\Gamma^{(0)}$, 
as a function of the dimensionless mass parameter $m$. The horizontal dashed line represents the massless limit, 
the solid line corresponds to the analytic result obtained through DREG, and the dots are obtained numerically through LTD unsubtraction. \label{fig:plotGammaTildesTOYcom}}
\end{center}
\end{figure}

Finally, we numerically integrate simultaneously the virtual and real corrections from Eqs.~(\ref{eq:toyV}),  (\ref{eq:GammaTilde1TOY}) and (\ref{eq:GammaTilde2TOY}) with the help of \Eq{eq:inverse} to obtain the total decay rate at NLO, $\Gamma^{(1)}$, as a function of 
the dimensionless mass parameter $m$. The result is shown in Fig.~\ref{fig:plotGammaTildesTOYcom}, and it is compared with 
the DREG analytic expression given by~\Eq{eq:TOYresultadoDREG}. The agreement is excellent and quite stable numerically. 
Moreover, the massless transition is very smooth because the momentum mappings are optimised to deal with the quasi-collinear 
configurations. In other words, the massless limit can directly be taken at the integrand level. 
This is another interesting advantage of the LTD approach.

\section{Unintregrated wave function and mass renormalisation for heavy quarks} 
\label{ssec:renormalisation}
In order to consider physical processes with heavy quarks we should also take into account self-energy corrections.
The well-known expressions of the wave function and mass renormalisation constants, 
in the Feynman gauge with on-shell renormalisation conditions, 
\bea
\Delta Z_2 &=& = \aas \, C_F \left(-\frac{1}{\ep_{\uv}}-\frac{2}{\ep_{\ir}}+ 3\,  \ln{\frac{M^2}{\mu^2}}-4\right)~, \nn
\\ \Delta Z_M^{\os} &=& \aas \, C_F \left(-\frac{3}{\ep_{\uv}}+ 3\,  \ln{\frac{M^2}{\mu^2}}-4\right)~,
\label{eq:waveandmassZ}
\eea
are not suitable, in particular, for the implementation of a local subtraction of the IR singularities. 
We shall provide unintegrated expressions. The case of massless quarks has been 
studied in detail in Ref.~\cite{Sborlini:2016gbr}. 
In~\Eq{eq:waveandmassZ}, we explicitly identify the origin of the $\ep$-poles, and the IR singularities 
of the wave function should cancel the IR singularities arising from the squared amplitudes of 
the real processes with radiated gluons off quarks. 

We consider  the process in which there are two on-shell massive fermions with momenta $p_1$ (quark) and $p_2$ (antiquark). 
The explicit one-loop self-energies are given by 
\bea
-\imath\Sigma(\bp_1)&=&\imath \, \gs^2 \, C_F \, \int_{\ell}\left(\prod_{i=1,3} G_F(q_i)\right)
\gamma^\mu \, (-\bq_3+M) \, \gamma^{\nu} \, d_{\mu\nu}(q_1)~, \\
-\imath\Sigma(-\bp_2)&=&\imath \, \gs^2 \, C_F \, \int_{\ell}
\left(\prod_{i=1,2} G_F(q_i)\right)\gamma^\mu(-\bq_2+M) \, \gamma^{\nu} \, d_{\mu\nu}(q_1)~.
\eea
In these expressions, we keep the same internal momenta $q_i$ that were used to define the vertex 
corrections in Sec.~\ref{sec:massivetriangle} (See also Fig.~\ref{fig:GenericNLObasicos} in Appendix~\ref{sec:physicalamplitudes}). 
This will allow us to reuse the same momentum mappings already defined to treat the vertex corrections. 
Because of the symmetry $\bp_1\leftrightarrow-\bp_2$, the unintegrated expression for the antiquark 
self-energy corrections can be deduced from those of the quark. 
Thus, we consider in the next only the quark self-energy. In the Feynman gauge
\bea
\Sigma(\bp_1) 
&=& \gs^2 \, C_F \, \int_{\ell} \left(\prod_{i=1,3} G_F(q_i)\right) ((d-2) \, \bq_3 + d \, M)~.
\label{eq:quarkself} 
\eea

Working in the \emph{on-shell renormalisation scheme} (${\rm OS}$), the renormalised self-energy fulfills
\beq
\Sigma_R(\bp_1=M) = 0 \, , \quad \quad \quad \quad \left.\frac{d\Sigma_R(\bp_1)}{d \bp_1}\right|_{\bp_1=M} =0~.
\label{eq:OSselfenergy}
\eeq
from where the wave function and mass renormalisation corrections are given by 
\beq
\Delta Z_2 = \left. \frac{\partial}{\partial \bp_1} \, \Sigma(\bp_1)  \, \right|_{\bp_1=M}~, \qquad
\Delta Z_M^{\os} = - \frac{1}{M} \, \Sigma(\bp_1=M)~.
\eeq
Explicitly, from \Eq{eq:quarkself} we obtain 
\beqn
\Delta Z_2(p_1) &=& -\g^2 \, C_F \, \int_{\ell} G_F(q_1) \, G_F(q_3) \, \left((d-2)\frac{q_1 \cdot p_2}{p_1 \cdot p_2}+4 M^2 \left(1- \frac{q_1 \cdot p_2}{p_1 \cdot p_2}\right)G_F(q_3)\right)~,
\label{eq:SEZ2expression}
\\ \Delta Z^{\os}_M(p_1) &=& -\g^2 \, C_F \, \int_{\ell} G_F(q_1) \, G_F(q_3) \, \left((d-2)\frac{q_1 \cdot p_2}{p_1 \cdot p_2}+2\right)~.
\label{eq:SEZMexpression}
\eeqn
It is worth to stress that the expression of the wave function renormalisation constant in \Eq{eq:SEZ2expression}
tends smoothly in the massless limit to the corresponding expression given in Ref.~\cite{Sborlini:2016gbr}. 
It is also relevant to notice that the term proportional to $M^2 (G_F(q_3))^2$ leads to soft divergences when $q_1$
gets on-shell. They are expected to cancel the soft divergences of the squared amplitudes of the real corrections. 
The dual representation of the mass renormalisation constant is straightforward from the LTD theorem.
The term $M^2 (G_F(q_3))^2$ in \Eq{eq:SEZ2expression}, however, introduces double poles that need to be 
treated specifically~\cite{Bierenbaum:2012th}. The final dual representations for both renormalisation 
factors are 
\bea
\Delta Z_2(p_1) &=&  \gs^2 \, C_F \, 
\int_\ell \Bigg[ \frac{\td{q_1}}{-2q_1\cdot p_1} \,  
\left(  (d-2) \,  \frac{q_1\cdot p_2}{p_1\cdot p_2} 
- \frac{4M^2}{2q_1\cdot p_1} \,  \left( 1 - \frac{q_1\cdot p_2}{p_1\cdot p_2} \right) \right) 
+ \frac{\td{q_3}}{2M^2+2q_3\cdot p_1}  \nn\\  &\times&  \bigg( (d-2)
\left( 1 + \frac{q_3\cdot p_2}{p_1\cdot p_{2}} \right)  
+ \frac{4M^2}{p_1\cdot p_2} \bigg(-\frac{\qb_3\cdot \pb_2}{2(q_{3,0}^{(+)})^2}  +
\frac{(q_{3,0}^{(+)}+p_{1,0}) \, q_3\cdot p_2}{q_{3,0}^{(+)}\, (2M^2+2q_3 \cdot p_1)} \bigg) \bigg) \Bigg]~, \nn \\
\Delta Z_M^{\os} (p_1) &=& \gs^2 \, C_F \,  \int_\ell \bigg[  \frac{\td{q_1}}{-2q_1\cdot p_1} \, 
\left( (d-2)\, \frac{q_1\cdot p_2}{p_1\cdot p_2} + 2  \right) 
+ \frac{\td{q_3}}{2M^2+2q_3\cdot p_1}  \left(  (d-2)\, \frac{q_3\cdot p_2}{p_1\cdot p_{2}} + d  \right) \bigg]~,  \nn \\
\eea
or in term of dual variables defined in Sec.~\ref{sec:massivetriangle} 
\bea
\Delta Z_2(p_1) &=& \gs^2 \, C_F \, \bigg[ \int 
\frac{2\, d[\xiu]\, d[v_1]}{1-\beta(1-2v_1)} \, \bigg( - (d-2) \, \frac{\xiu \, (1+\beta(1-2v_1))}{1+\beta^2}  \nn \\
&+& \frac{2m^2}{1-\beta(1-2v_1)} \, \left(\frac{1}{\xiu}-\frac{1+\beta(1-2v_1)}{1+\beta^2} \right) \bigg) \nn \\
&+& \int \frac{2 \xi_3^2 \, d[\xi_3] \, d[v_3]}{\xit\, (\xit - \beta\, \xi_3 (1-2v_3) + m^2)}
\bigg(  (d-2) \, \left( 1 + \frac{\xit +\beta \, \xi_3 \, (1-2v_3)}{1+\beta^2} \right) \nn \\ 
&+& \frac{2m^2}{(1+\beta^2) \, \xit} \left( \frac{\beta\, \xi_3 \, (1-2v_3)}{\xit}
+ \frac{(1+\xit)(\xit + \beta\, \xi_3 \, (1-2v_3))}{\xit - \beta\, \xi_3 (1-2v_3) + m^2} \right) \bigg)~, \nn \\ 
\Delta Z^{\os}_M(p_1) &=& \gs^2 \, C_F \, \bigg[ -\int  
\frac{2\, d[\xiu]\, d[v_1]}{1-\beta(1-2v_1)} \, \left((d-2) \, \frac{\xiu \, (1+\beta(1-2v_1))}{1+\beta^2} +2 \right) \nn \\ 
&+& \int \frac{2 \, \xi_3^2\, d[\xi_3]\, d[v_3]}{\xit (\xit - \beta \, \xi_3 \, (1-2v_3) + m^2)}
\left( (d-2) \, \frac{\xit +\beta \, \xi_3 \, (1-2v_3)}{1+\beta^2}+ d\right) \bigg]~. 
\eea

\section{UV renormalisation}
\label{sec:renormalisation}

We shall now remove the UV divergences of the renormalisation constants by defining suitable integrand level 
UV counter-terms. These UV counter-terms are obtained by expanding Eqs. (\ref{eq:SEZ2expression}) and (\ref{eq:SEZMexpression}) 
around the UV propagator $G_F(q_{\uv}) = 1/(q_{\uv}^2-\mu_{\uv}^2+i0)$,  
where $q_{\uv} = \ell + k_{\uv}$ with $k_\uv$ arbitrary~\cite{Becker:2010ng,Hernandez-Pinto:2015ysa,Sborlini:2016gbr}. 
The simplest choice is $k_\uv =0$. We obtain 
\beqn
\Delta Z_2^{\uv}(p_1) &=& -(d-2)\, \g^2 \, C_F \, \, \int_{\ell} (G_F(q_\uv))^2 \, \left(1+\frac{q_\uv \cdot p_2}{p_1 \cdot p_2}\right) \nn \\ 
&\times& \, \left(1- G_F(q_\uv)(2\, q_\uv \cdot p_1 + \mu^2_\uv)\right)~, \nn \\
\Delta Z_M^{\os, \uv}(p_1) &=& - \g^2 \, C_F \, \, \int_{\ell} (G_F(q_\uv))^2 \, \left(d+(d-2)\frac{q_\uv \cdot p_2}{p_1 \cdot p_2}\right) \nn \\
&\times& \, \left(1- G_F(q_\uv)(2 q_\uv \cdot p_1 + 2 d^{-1} \mu^2_\uv)\right)~,
\label{eq:zetasUV}
\eeqn
whose integrated form is
\bea
\nn && \Delta Z_2^{\uv} = - \Se \, \aas \, C_F \,  \left( \frac{\mu_\uv^2}{\mu^2}\right)^{-\ep} \, \frac{1-\ep^2}{\ep}~,
\label{eq:Z2uvintegrado}
\\ && \Delta Z_M^{\os, \uv} = - \Se \, \aas \, C_F \,  \left( \frac{\mu_\uv^2}{\mu^2}\right)^{-\ep} \, \frac{3}{\ep}~.
\label{eq:ZMuvintegrado}
\eea
The sub-leading terms in~\Eq{eq:zetasUV}, which are proportional to $\mu_\uv^2$,
have been adjusted in such a way that only the UV poles in \Eq{eq:waveandmassZ} are 
subtracted at ${\cal O}(\ep^0)$. Therefore
\beq
\Delta Z_2^{\ir} = \Delta Z_2 - \Delta Z_2^{\uv}~, \qquad 
\Delta Z_M^{\os, \ir} =  \Delta Z_M^{\os}  - \Delta Z_M^{\os, \uv}~,
\eeq
only contain IR singularities, including the finite terms which are scheme dependent.

The dual representation of \Eq{eq:zetasUV} requires to evaluate the residue of poles of second and 
third order~\cite{Bierenbaum:2012th,Hernandez-Pinto:2015ysa,Sborlini:2016gbr} located at
$q^{(+)}_{\uv,0}=\sqrt{\textbf{q}_{\uv}^2+\mu_{\uv}^2-\imath 0}$. We obtain 
\bea
\Delta Z_2^{\uv} &=& - (d-2) \, \gs^2 \, C_F \, \int_\ell \frac{\td{q_\uv}}{2 \left( q_{\uv,0}^{(+)} \right)^2} \, 
\bigg[ \left(1 - \frac{\qb_\uv \cdot \pb_{2}}{p_1\cdot p_{2}} \right)\, \nn \\ &\times&
\bigg( 1- \frac{3 ( 2 \qb_\uv \cdot \pb_1 - \mu_\uv^2)}{4 \left( q_{\uv,0}^{(+)} \right)^2}\bigg) 
-\frac{p_{1,0} \, p_{2,0}}{2 p_1\cdot p_2} \bigg]~, \nn \\
\Delta Z_M^{\os, \uv} &=& - \gs^2 \, C_F \, \int_\ell \frac{\td{q_\uv}}{2 \left( q_{\uv,0}^{(+)} \right)^2} \,
\bigg[ \left(d - (d-2)\frac{\qb_\uv \cdot \pb_{2}}{p_1\cdot p_{2}} \right)\, \nn \\ &\times&
\bigg( 1- \frac{3 ( 2 \qb_\uv \cdot \pb_1 - 2 d^{-1} \mu_\uv^2)}{4 \left( q_{\uv,0}^{(+)} \right)^2}\bigg)  
- (d-2)\, \frac{p_{1,0} \, p_{2,0}}{2 p_1\cdot p_2} \bigg]~.
\eea
Then, we use the parametrisation
\beqn
\nn && q_\uv^{\mu} = \frac{\sqrt{s_{12}}}{2} \, 
\left(\xi_{\uv,0}, 2 \, \xi_{\uv} \, \sqrt{v_\uv(1-v_\uv)} \, {\bf e}_{\uv, \perp}, \xi_{\uv} \, (1-2v_\uv) \right)~, 
\\ && \xi_{\uv,0} = \sqrt{m_{\uv}^2+\xi_{\uv}^2}~,
\eeqn
with $m_{\uv} = 2 \mu_{\uv}/\sqrt{s_{12}}$, and
the UV counter-term get the form 
\bea
\Delta Z_2^{\uv}  &=& - (d-2) \, \gs^2 \,  C_F \, \int d[\xiuv] \, d[v_{\uv}] \, 
\frac{2\xiuv^2}{\xi_{\uv,0}^3} \bigg[ \bigg(1 + \frac{\beta \, \xiuv\, (1-2v_{\uv})}{2(1+\beta^2)}\bigg) \nn \\ &\times& 
\bigg( 1 - \frac{3 \, \left( 2 \, \beta\, \xiuv \, (1-2v_{\uv}) - m^2_{\uv} \right)}
{4\, \xi_{\uv,0}^2} \bigg) - \frac {1}{2(1+\beta^2)} \bigg]~, \nn \\
\Delta Z_M^{\os, \uv}  &=& -  \gs^2 \,  C_F \, \int d[\xiuv] \, d[v_{\uv}] \, 
\frac{2\xiuv^2}{\xi_{\uv,0}^3} \bigg[ \bigg(d + (d-2)\, \frac{\beta \, \xiuv\, (1-2v_{\uv})}{2(1+\beta^2)}\bigg) \nn \\ &\times& 
\bigg( 1 - \frac{3 \, \left( 2 \, \beta\, \xiuv \, (1-2v_{\uv}) - 2 d^{-1}\, m^2_{\uv} \right)}
{4\, \xi_{\uv,0}^2} \bigg) - \frac {d-2}{2(1+\beta^2)} \bigg]~.
\label{eq:DZetaUV}
\eea

Similarly, we should subtract the UV singularities of the $q\bar q A$ interaction vertex, 
with $A=\{\phi, \gamma, Z\}$ for the explicit examples that we will consider later.
As for the self-energy contributions, the UV counter-term is obtained by expanding the vertex corrections around the UV propagator  
$G_F(q_{\uv}) = 1/(q_{\uv}^2-\mu_{\uv}^2+i0)$. In the Feynman gauge, the generic expression of the vertex UV counter-term reads 
\beq
{\bf \Gamma}^{(1)}_{A, \uv} = \gs^2 \, C_F\, 
\int_\ell \left( G_F(q_\uv)\right)^3 \, 
\left[\gamma^{\nu} \, \bq_{\uv} \, {\bf \Gamma}^{(0)}_A \, \bq_{\uv} \, \gamma_{\nu} - 
d_{A, \uv} \, \mu_{\uv}^2 \, {\bf \Gamma}^{(0)}_{A}\right]~,
\label{eq:genericvertex}
\eeq 
where the the tree-level vertices ${\bf \Gamma}^{(0)}_{A}$ are given in \Eq{eq:treelevelvertices} of Appendix~\ref{sec:physicalamplitudes}.
The $\mu_{\uv}^2$ term is sub-leading and the coefficient $d_\uv$ is adjusted to subtract only the UV pole. 
Performing the explicit calculation, we find that in the $\overline{{\rm MS}}$ scheme these coefficients are
\beq
d_{\phi,\uv} = d+4~, \qquad d_{\gamma,\uv} = d_{Z,\uv} = d~.
\eeq
Notice that this choice of the sub-leading contributions differs
from $d_{\gamma, \uv}=4$ proposed in Ref.~\cite{Becker:2010ng}. 
The difference is, however, of ${\cal O}(\ep)$. 
The integration of the vertex UV counter-term leads to the result 
\beq
{\bf \Gamma}^{(1)}_{A, \uv} = \widetilde{S}_{\ep} \,  \aas \, C_F\, 
{\bf \Gamma}^{(0)}_{A} \, \left( \frac{\mu_\uv^2}{\mu^2}\right)^{-\ep} \frac{c_{A,\uv}}{\ep} 
\eeq
with 
\beq
c_{\phi,\uv} = 4~, \qquad c_{\gamma,\uv} = c_{Z,\uv} = 1~.
\label{eq:ccoefficients}
\eeq
that translates into 
\beq
\la \M{0}_{A}|\M{1}_{A, \uv} \ra = \widetilde{S}_{\ep} \,  \aas \, C_F\,
|\M{0}_A|^2 \, \left( \frac{\mu_\uv^2}{\mu^2}\right)^{-\ep} \frac{c_{A,\uv}}{\ep}~. 
\eeq

The dual representation of \Eq{eq:genericvertex} is (see Ref.\cite{Sborlini:2016gbr})
\bea
{\bf \Gamma}^{(1)}_{A, \uv} &=& \gs^2 \, C_F\, 
\int_\ell \frac{\td{q_\uv}}{8 \left(q_{\uv,0}^{(+)}\right)^2}   \bigg[
\gamma^{\nu} \, \gamma^0 \, {\bf \Gamma}^{(0)}_A \, \gamma^0 \, \gamma_{\nu} \nn \\ 
&-&\frac{3}{\left(q_{\uv,0}^{(+)}\right)^2}
\left[\gamma^{\nu} \,  (\gamma\cdot \qb_{\uv}) \, {\bf \Gamma}^{(0)}_A \, (\gamma\cdot \qb_{\uv}) \, \gamma_{\nu} - 
d_{A, \uv} \, \mu_{\uv}^2 \, {\bf \Gamma}^{(0)}_{A}\right]~.
\eea
After an explicit calculation, we obtain for the vertex UV counter-terms 
\bea
\la \M{0}_{\phi} | \M{1}_{\phi,\uv} \ra &=& \gs^2 \, C_F\, |\M{0}_{\phi}|^2 \, \int d[\xi_{\uv}]\, d[v_{\uv}] \, \frac{2\, \xiuv^2}{\xi_{\uv,0}^3}\, 
\left( 7-2\ep - \frac{3\xiuv^2}{\xi_{\uv,0}^2} \right)~, \nn \\
\la \M{0}_{\gamma} | \M{1}_{\gamma,\uv}\ra &=& \gs^2 \, C_F\, \int d[\xi_{\uv}]\, d[v_{\uv}] \, \bigg[
\frac{2\, (e\, e_q)^2\, C_A}{1-\ep} \, f(\xi_{\uv},v_{\uv}) \nn \\ &+& |\M{0}_{\gamma}|^2 \, 
\frac{\xiuv^2}{\xi_{\uv,0}^3}\, \bigg( 7-4\ep - \frac{3\xiuv^2}{\xi_{\uv,0}^2} (1+4v_\uv(1-v_\uv))\bigg) \bigg]~, \nn \\ 
\la \M{0}_{Z} | \M{1}_{Z,\uv}\ra &=& \gs^2 \, C_F\, \int d[\xi_{\uv}]\, d[v_{\uv}] \, \bigg[ 
\frac{2\, g_{V,q}^2\, C_A}{1-\ep}\, f(\xi_{\uv},v_{\uv}) \nn \\ &+&  |\M{0}_{Z}|^2 \, 
\frac{\xiuv^2}{\xi_{\uv,0}^3}\, \bigg( 7-4\ep - \frac{3\xiuv^2}{\xi_{\uv,0}^2} (1+4v_\uv(1-v_\uv))\bigg) \bigg]~, 
\label{eq:UVunintegrated}
\eea
where the function 
\beq
f(\xi_{\uv},v_{\uv}) = 24 M^2 \, \frac{\xiuv^4}{\xi_{\uv,0}^5}  \left(\ep (1-2v_{\uv})^2 + 6v_{\uv}(1-v_{\uv})-1\right)~,
\eeq
integrates to zero and does not contribute to the renormalisation of the vertex. 
However, this additional term is necessary to achieve a local cancellation of the UV behaviour. 

The UV divergences of the wave function cancel exactly the UV divergences of the vertex corrections
for photons and $Z$ bosons, because conserved currents or partially conserved currents, as the
vector and axial ones, do not get renormalised. 
The corresponding dual representations, however, do not cancel each other at 
integrand level. In particular, the wave function renormalisation contains linear UV singularities that 
cancel upon integration. Also, the vertex UV counter-term contains terms that are proportional to the mass and
cancel upon integration. The contribution of all this spurious terms is, however, crucial to cancel locally all the UV 
singularities. The coupling to scalar particles, on the contrary, needs to be 
renormalised
\beq
Y_q^0 \, \mu_0^\ep = Y_q \, \mu^\ep \left(1 -\aas \frac{3C_F}{\ep} \right) + {\cal O}(\as^2)~.
\eeq

\section{LTD four-dimensional unsubtraction for physical processes}
\label{sec:physicalexamples}
\begin{figure}[t]
\begin{center}
\includegraphics[width=0.45\textwidth]{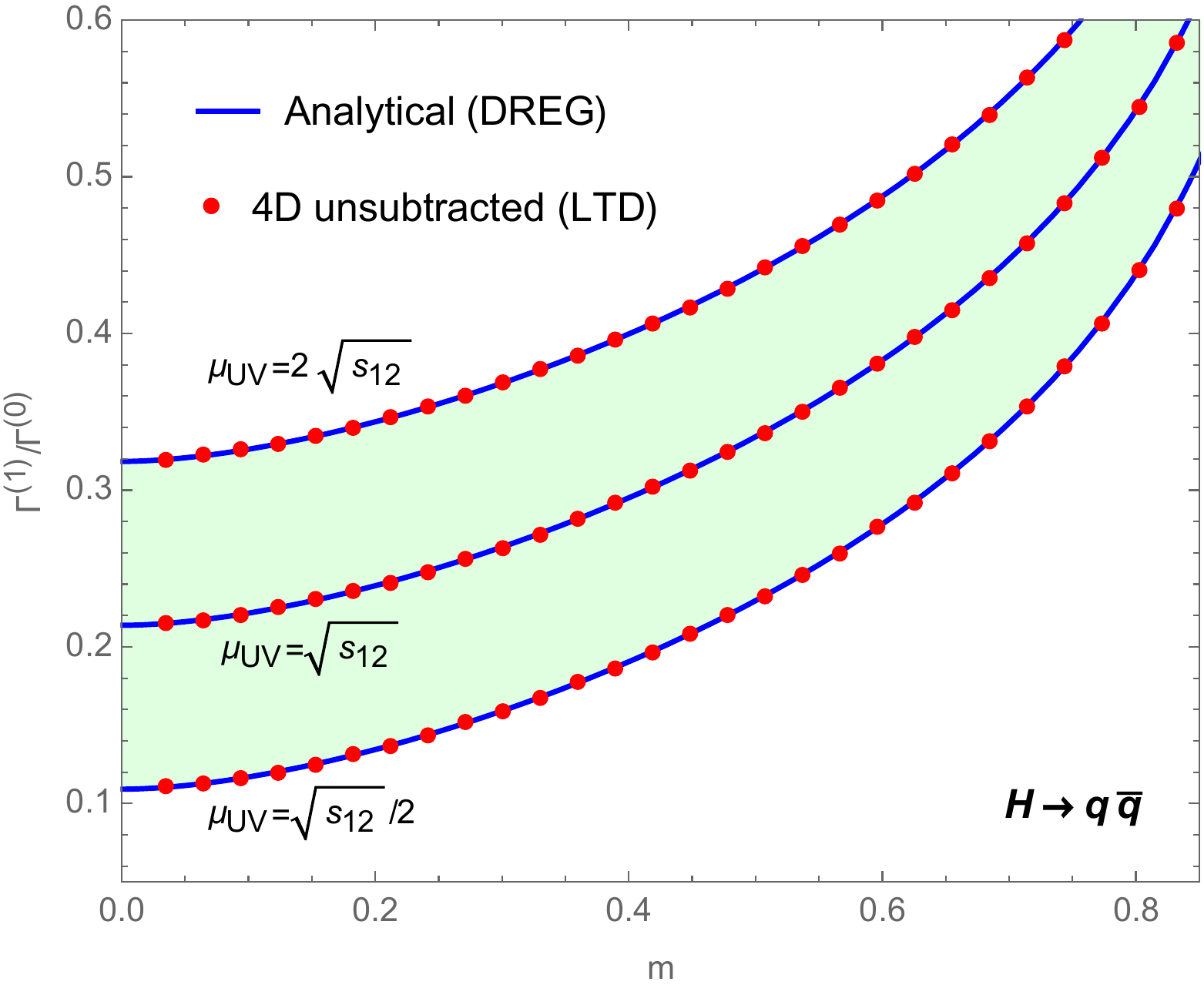} \qquad
\includegraphics[width=0.45\textwidth]{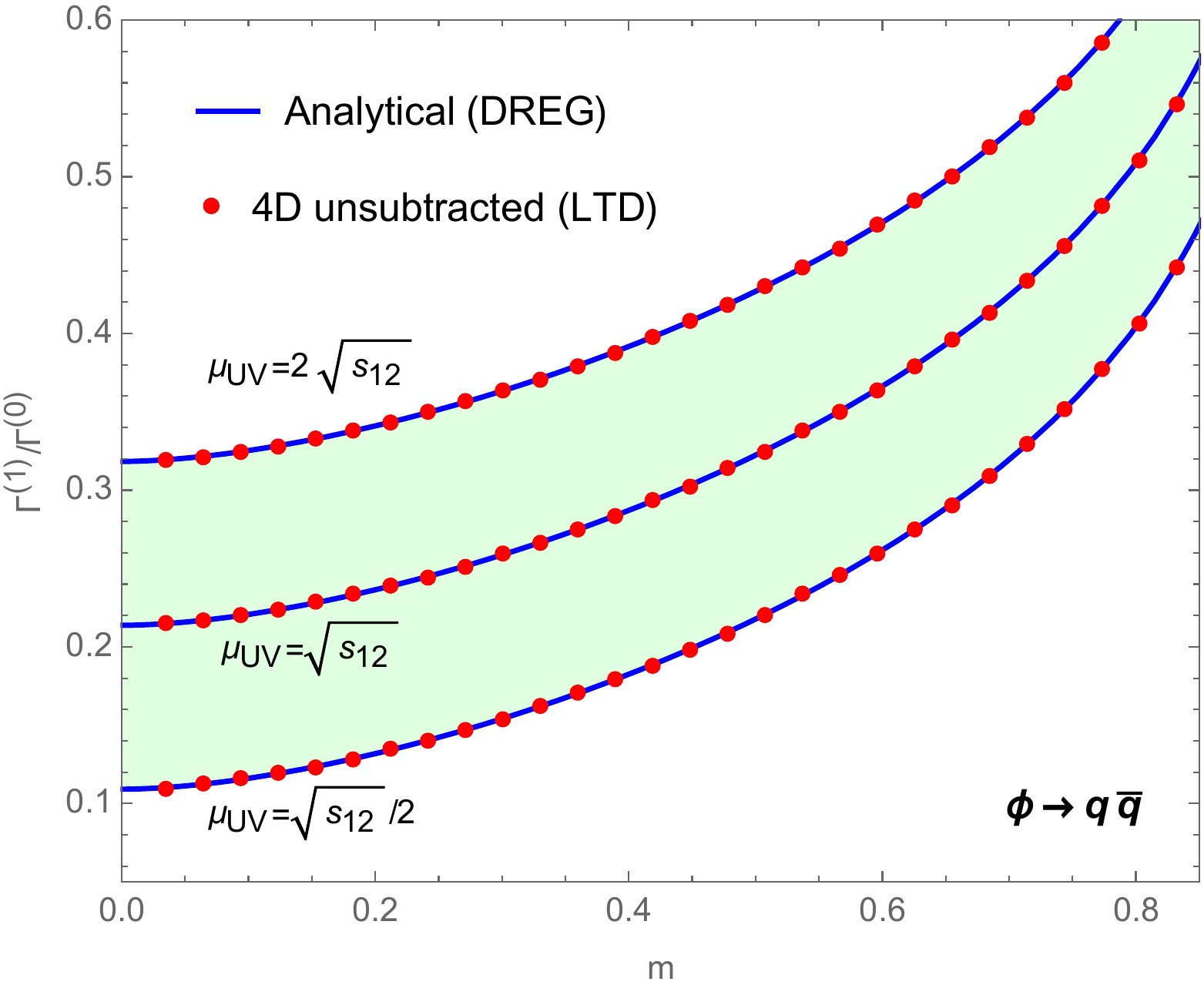}
\caption{Total decay rate at NLO for scalar and pseudoscalar particles into a pair of heavy quarks as a function of the mass,
normalised to the LO. In the left panel, we consider a standard Higgs boson, whilst in the right panel 
we plot the decay rate for a pseudoscalar particle ($c_q=1/2$). 
The solid blue lines correspond to the usual DREG analytic result, while the red dots were computed numerically 
within the LTD unsubtracted method. We also consider a renormalisation scale 
variation, in the range $1 / 2 < \mu_{\uv}/\sqrt{s_{12}} <2$. 
\label{fig:LTDHiggses}}
\end{center}
\end{figure}

\begin{figure}[t]
\begin{center}
\includegraphics[width=0.6\textwidth]{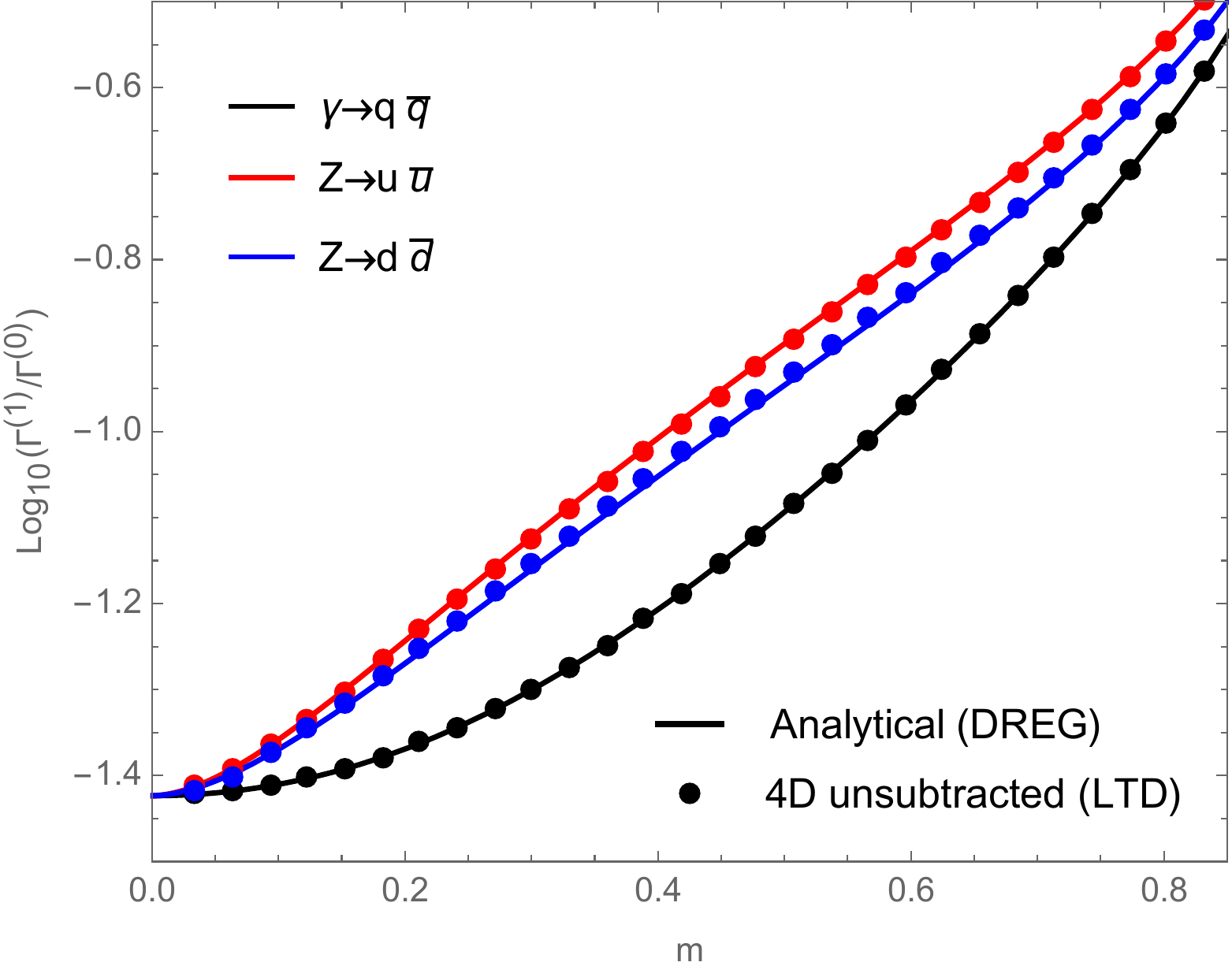}
\caption{ Total decay rate at NLO for off-shell vector particles into a pair of heavy quarks as a function of the mass, 
normalised to the LO. We consider three physical cases: $\gamma^* \to q \bar q$ (black), 
$Z^ * \to u \bar u$ (red, up-type quarks) and $Z^ * \to d \bar d$ (blue, down-type quarks). 
Solid lines corresponds to the results within the DREG approach, 
whilst the dots we obtained with the LTD unsubtracted method. 
\label{fig:LTDVectores}}
\end{center}
\end{figure}

We have already defined all the necessary ingredients to test the four dimensional implementation 
of NLO corrections to physical processes in the LTD framework. In particular, 
we will compute the NLO QCD corrections to the decay rate $A^* \to q \bar q(g)$, with $A=\phi, \gamma, Z$. 
The actual implementation is indeed independent of the decaying particle. 
The renormalised one-loop amplitude is given by 
\beq
|\M{1,\r}_A\ra =  |\M{1}_A\ra - |\M{1,\uv}_A \ra + \frac{1}{2} \left( \Delta Z_2^{\ir}(p_1) + \Delta Z_2^{\ir}(p_2) \right) |\M{0}_A\ra~, 
\eeq
where $|\M{1,\uv}_A \ra$ is the unintegrated UV counter-term of the one-loop vertex correction, $|\M{1}_A\ra$, and $Z_2^{\ir}(p_i)$ are the 
IR components of the quark and antiquark self-energy corrections. From the renormalised one-loop amplitude $|\M{1,\r}_A\ra$, which  
contains only IR singularities, we construct the LTD representation of the renormalised virtual decay rate 
\beqn
\Gamma_{\v,A}^{(1,\r)} &=& \frac{1}{2\, \sqrt{s_{12}}} \, 
\sum_{i=1}^3 \int d\Phi_{1\to 2} \, 2 {\rm Re} \la {\cal M}_A^{(0)} | {\cal M}_A^{(1,\r)}(\td{q_i}) \ra~.
\label{eq:dualvirtualPHYS}
\eeqn
The corresponding dual amplitudes for the vertex corrections are given explicitly
in Eqs.~(\ref{eq:phidual}), (\ref{eq:gammadual}) and (\ref{eq:Zdual}).
As for the toy scalar example presented in Sec.~\ref{sec:unsubtraction}, the real contributions are implemented 
by splitting the real phase-space in two domains 
\beq
\widetilde{\Gamma}^{(1)}_{\r,A,i} = \frac{1}{2 \sqrt{s_{12}}} \, \int\, d\Phi_{1\to 3} \, | \M{0}_{A\to \qq g}|^2 \, 
{\cal R}_i\left(y'_{ir} < y'_{jr}\right) \, , \qquad i,j=\{1,2\}~,
\label{eq:dualrealPHYS}
\eeq 
with $\Gamma^{(1)}_{\r,A}=\widetilde{\Gamma}^{(1)}_{\r,A,1}+\widetilde{\Gamma}^{(1)}_{\r,A,2}$ the real total decay rate.
The real emission squared amplitudes are given in \Eq{eq:Aqqbarg}.
In each of the real-phase space domain we introduce one of the momentum mappings defined in Sec.~\ref{sec:realvirtualmap}. 
The sum of the virtual and real corrections in \Eq{eq:dualvirtualPHYS} and \Eq{eq:dualrealPHYS} is a single integral in 
the loop three-momentum.  It is UV and IR finite locally and thus can be calculated numerically with $\ep=0$.
We follow the same numerical implementation as for the scalar example presented in Sec.~\ref{sec:unsubtraction}, and 
compare the numerical output with the analytical total decay rate, which has the form 
\beqn
\Gamma^{(1)}_A &=& \aas \,  C_F \, \left[ \Gamma_{A}^{(0)} \, \left( F(x_S) +  
2 (c_{A,\uv}-1) \, \ln{\frac{\mu_\uv^2}{s_{12}}}\right) + G_{A}(x_S) \right] \, + {\cal O}(\ep)~.
\label{eq:SUMArealvirtualRENORMALIZED}
\eeqn
Our results, normalised to the LO decay rate $\Gamma^{(0)}_A$,  
are presented in Fig.~\ref{fig:LTDHiggses} for a scalar and pseudoscalar, and
in Fig.~\ref{fig:LTDVectores} for vector bosons. The agreement with the analytic prediction 
is excellent in all the cases. Moreover, the massless limit, i.e. $x_S \to 0$, is also 
well-defined and we recover the known results. This is a very subtle point, because individual 
contributions in DREG are not smoothly well-defined in that limit.

\section{Conclusions and outlook}
\label{sec:conclusions}
In this article, we have generalised the four-dimensional unsubtraction 
method~\cite{Hernandez-Pinto:2015ysa,Sborlini:2015uia,Sborlini:2016fcj,Sborlini:2016gbr,ll2016} 
to deal with massive particles. Based in the LTD theorem, it exploits the possibility of expressing virtual amplitudes 
as phase-space integrals, and the fact that threshold and IR singularities are always restricted to a compact region 
of the loop three-momentum integration domain. This is a crucial point, because it allows to establish a momentum 
mapping to generate the real-emission on-shell kinematics starting from the Born level momenta and 
the loop three-momentum, in such a way that a local cancellation of both IR and UV singularities 
can be achieved without introducing IR subtractions. In particular for the massive case, we have defined 
a general momentum mapping that accounts properly for the quasi-collinear configurations.  

First, we started by inquiring in the computation of the scalar three-point function with massive 
particles within the LTD approach. Besides recovering the previously known results, the analysis of the 
integration domains of the dual contributions allowed us to understand the origin of its singular structure. 
Then, we illustrated the local cancellation of IR and quasi-collinear configurations with a toy scalar example. 

The full cancellation of IR singularities requires the contributions of the self-energy corrections. 
Thus, we defined unintegrated versions of the quark wave function and mass renormalisation factors 
in the on-shell renormalisation (OS) scheme.  Compared to the massless case, this is a non-trivial case 
because the OS scheme is built at integral level and DREG leads to very 
simple results after integration. In any case, the unintegrated renormalisation constants are 
completely general and lead to a fully local cancellation of the remaining IR singularities. 
The treatment of UV singularities was also discussed carefully, and we constructed suitable 
unintegrated UV counter-terms for both self-energy and vertex corrections, which reproduce 
successfully the $\overline{\rm MS}$ conditions and also lead to a fully local cancellation of UV singularities.  

Finally, we tested the LTD four-dimensional unsubtraction to compute NLO QCD corrections 
to the decay rate of scalar and vector particles into a pair of massive quarks. The results were 
compared with the standard DREG expressions, and we found an impressive agreement. 
In particular, the transition to the massless limit is smooth because the quasi-collinear 
configurations of the real and virtual corrections are matched at integrand level. 
With the results presented in this paper, LTD four-dimensional unsubtration 
can be applied to any multipartonic process involving heavy quarks, and other 
heavy particles. 

\section*{Acknowledgements}
This work has been supported by CONICET Argentina,
by the Spanish Government and ERDF funds from the European Commission 
(Grants No. FPA2014-53631-C2-1-P and SEV-2014-0398)
and by Generalitat Valenciana under Grant No. PROMETEOII/2013/007. 
FDM acknowledges suport from Generalitat Valenciana (GRISOLIA/2015/035).

\appendix

\section{Phase-space}
\label{app:PSformulae}
The phase-space for a $1 \to 2$ decay with final-state particles of equal masses,
$p_i^2=M^2$ with $i=1,2$ and $s_{12}$ the virtuality of the decaying particle, 
is given by 
\beq
\int d\Phi_{1\to 2} = \frac{\Gamma(1-\ep) \, \beta^{1-2\ep}}{2(4\pi)^{1-\ep}\, \Gamma(2-2\ep)}
\left( \frac{s_{12}}{\mu^2}\right)^{-\ep}~.
\eeq
with $\beta = \sqrt{1-m^2}$ and $m^2 = 4M^2/s_{12}$.
The corresponding $1 \to 3$ phase-space of the real radiation correction with an additional massless particle in the final state, 
$(p'_r)^2=0$, is given by
\beq 
d\Phi_{1\to 3} = \frac{(4\pi)^{\ep-2}\, s_{12}}{\Gamma(1-\ep)} \left( \frac{s_{12}}{\mu^2}\right)^{-\ep} 
\left( \int d\Phi_{1\to 2} \right) \, \beta^{-1+2\ep} \, \theta(h_p) \, h_p^{-\ep} \, dy_{1r}'  \, dy_{2r}'~, 
\label{eq:PHASESPACE}
\eeq
with 
\beq
h_p = (1 - y_{1r}' - y_{2r}' )\, y_{1r}' \, y_{2r}' - \frac{m^2}{4} (y_{1r}'+y_{2r}')^2~,
\label{eq:hp}
\eeq
where $y_{ir}'=2\, p_i'\cdot p_r' /s_{12}$. 
In order to integrate analitically the real radiation contribution, it is convenient to use the change of 
variables suggested in Ref. \cite{Bilenky:1994ad}, i.e.
\beqn
&& y_{1r}' = g(z) \, w~, \qquad y_{2r}'= g(z) \, z \, w~,
\label{eq:wzdef} \qquad 
g(z) = \frac{(z-x_S)\, (1-x_S\, z)}{z\, (1+z) \, (1+x_S)^2},
\label{eq:gzdef}
\eeqn
which allows to factorise the function $h_p$ in \Eq{eq:hp} according to
\beq
h_p = g(z)^3 \, z(1+z) \, w^2(1-w)~.
\eeq
In consequence, the phase-space limits, which are determined by the quadratic function $h_p$, simplify to $z \in [x_S,x_S^{-1}]$ and $w \in [0,1]$. 
The first integral in $w$ can easily be obtained by keeping the exact $\ep$-dependence. 
The second integral in $z$, however, requires to expand the result up to ${\cal O}(\ep^0)$ before integration.

\section{Unification of coordinates}
\label{app:unificationcoordinates}
In order to avoid local mismatches in the $\ep$-expansion close to the singular regions, it is necessary to unify the coordinate system and express all the on-shell momenta $q_i$ in terms of a single loop three-momentum~\cite{Hernandez-Pinto:2015ysa}. This is a crucial point to obtain integrable expressions directly at integrand level in four space-time dimensions . Since $q_3=\ell$, we parametrise the on-shell momenta $q_1$ and $q_2$ by using the variables $(\xi,v)\equiv (\xi_{3},v_3)$. Notice that each $q_i$ is set on-shell inside the corresponding dual contribution. Hence, by using the associated dispersion relations, we work with the spatial components of the momenta and fix the energy to fulfill the on-shell condition. From \Eq{Trianguloqiparametrizacion}, the spatial components of the loop momenta must satisfy
\beeq
\nn {\bf q}_1&=&\frac{\sqrt{s_{12}}}{2}\xi_{1,0}\left(2  \, \sqrt{v_1(1-v_1)} \, {\bf e}_{1,\perp}, 1-2v_1 \right)\\
&=&{\bf q}_3+{\bf p}_1= \frac{\sqrt{s_{12}}}{2} \, 
\left(2 \, \xi_{3} \, \sqrt{v_3(1-v_3)} \, {\bf e}_{3, \perp}, \xi_{3} \, (1-2v_3) + \beta\right)\,,
\eeeq
which leads to a system of equations, whose solution is
\beeq
\xi_{1,0}&=&\sqrt{(\beta+\xi)^2-4\beta \, v \, \xi}~,
\label{eq:ReemplazoXI10}
\\ v_1&=&\frac{1}{2}\left(1-\frac{\beta+(1-2v)\, \xi}{\xi_{1,0}}\right).
\label{eq:ReemplazoV1}
\eeeq
Similarly, we can express $(\xi_2, v_2)$ in terms of $(\xi,v)$ from $\qb_2=\qb_3$. This leads to the trivial replacement $(\xi_2,v_2) \to (\xi,v)$. It is worth appreciating that this change of reference frame is well defined because the argument of the square root in \Eq{eq:ReemplazoXI10} is always positive. In fact,
\beq
(\xi+\beta)^2-4 \beta \, \xi \, v >(\xi+\beta)^2-4\xi \, \beta = (\xi-\beta)^2 >0 \, ,
\eeq
due to $\beta<1$. Then, the associated Jacobian is given by
\beq
{\cal J}(\xi,v) = \frac{\xi ^2}{(\xi+\beta)^2-4 v \beta \xi} \, ,
\eeq
whose massless limit (i.e. $\beta \to 1$) agrees with the expressions found in Refs. \cite{Hernandez-Pinto:2015ysa,Sborlini:2016gbr}.

\section{LTD amplitudes for $A^* \to q \bar q (g)$}
\label{sec:physicalamplitudes}
\begin{figure}[t]
\begin{center}
\includegraphics[width=0.95\textwidth]{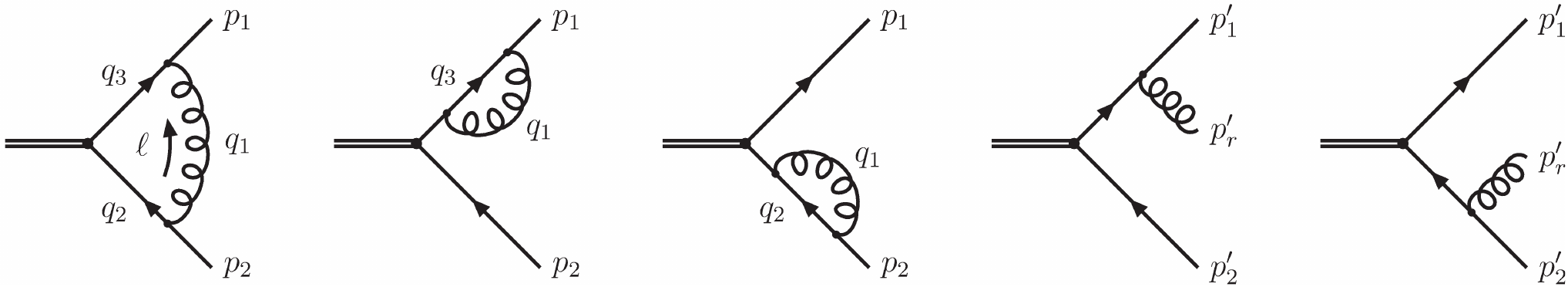}
\caption{Momentum configuration of the NLO QCD corrections to the process $A^* \to q\bar q (g)$, under the assumption that the 
decaying particle does not couple to gluons. 
\label{fig:GenericNLObasicos}}
\end{center}
\end{figure}

In this Appendix, we collect the dual amplitudes and real squared amplitudes contributing to the  
NLO QCD corrections to the processes $A^* \to q \bar q(g)$, with $A=\{\phi, \gamma, Z\}$. 
The tree-level vertices are given by
\beqn
&& {\bf \Gamma}^{(0)}_{\phi}= \imath \, Y_q \, \left(1+c_{q} \, \gamma^{5}\right)~,  \nn \\ 
&& {\bf \Gamma}^{(0)}_{\gamma} = \imath \, e \, e_q \, \gamma^{\mu}~, \nn \\
&& {\bf \Gamma}^{(0)}_{Z} = \imath \, \gamma^{\mu} \, \left(g_{V,q} + g_{A,q} \, \gamma^{5}\right)~,
\label{eq:treelevelvertices}
\eeqn
The corresponding Born squared amplitudes, averaged over the initial-state polarisations, are
\bea
|\M{0}_{\phi \to q \bar q}|^2 &=& 2 \, s_{12} \, Y_q^2 \, C_A \, \left(\beta^2 + c_q^2\right)~,
\\ |\M{0}_{\gamma \to q \bar q}|^2 &=& 2\, s_{12} \, (e\,  e_q)^2 \, C_A \, \left(1+\frac{m^2}{2(1-\ep)}\right)~,
\\ |\M{0}_{Z \to q \bar q}|^2 &=& 2\, s_{12} \, C_A \, \frac{2(1-\ep)}{3-2\ep} \left( g_{V,q}^2 \, \left(1+\frac{m^2}{2(1-\ep)}\right)
+ g_{A,q}^2 \,  \beta^2 \right)~.
\label{eq:bornsquared}
\eea
The momentum configuration of the NLO QCD corrections is represented in Fig.~\ref{fig:GenericNLObasicos}.

The squared  amplitudes for the real process $A^* \to q(p_1')+\bar{q}(p_2') + g(p_r')$, are given by 
\bea
\nn | \M{0}_{\phi \to \qq g}|^2 &=& 4 \, \gs^2\, C_F \, 
\bigg[ s_{12}^{-1} \, |\M{0}_{\phi \to \qq}|^2  \, \frac{h_p}{2\, (y_{1r}' \, y_{2r}')^2}  \nn \\
&+& Y_q^2 \left(1+c_q^2\right) \, C_A (1-\ep) \left(1+\frac{y_{2r}'}{y_{1r}'}\right) \bigg]
+ \, (1 \leftrightarrow 2), \nn \\
\nn |\M{0}_{\gamma \to \qq g}|^2 &=& 4 \, \gs^2 \, C_F \, 
\bigg[ s_{12}^{-1}\, |\M{0}_{\gamma \to \qq}|^2 \, \frac{h_p}{2\, (y_{1r}' \, y_{2r}')^2}
+(e \, e_q)^2 \, C_A \,  \left((1-\ep)\, \frac{y_{2r}'}{y_{1r}'}-\ep\right) \bigg] + \, (1 \leftrightarrow 2)~,
\\ \nn |\M{0}_{Z \to \qq g}|^2 &=& 4\, \gs^2\, C_F \, 
\bigg[ s_{12}^{-1}\, |\M{0}_{Z \to \qq}|^2 \, \frac{h_p}{2\, (y_{1r}' \, y_{2r}')^2} 
+ C_A \, \frac{2(1-\ep)}{3-2\ep} \nn \\ &\times& \left( (g_{V,q}^2+g_{A,q}^2)\, \left((1-\ep)\, \frac{y_{2r}'}{y_{1r}'}-\ep \right) 
+ g_{A,q}^2 \, \frac{m^2}{2} \left(1+\frac{y_{2r}'}{y_{1r}'} \right)\right) \bigg]  + \, (1 \leftrightarrow 2) \, ,
\label{eq:Aqqbarg}
\eea 
where the function $h_p$ is defined in~\Eq{eq:hp}. 

The vertex corrections to the process $A^*\to \qq$ are given by the dual amplitudes 
\bea
\la \M{0}_A|\M{1}_A(\td{q_1})\ra&=& \gs^2 \, C_F \, \int_{\ell} \td{q_1} \bigg[
- \frac{ |\M{0}_A |^2\, s_{12} (1+\beta^2)}{(2 q_1\cdot p_1)\,(2 q_1\cdot p_2)} +  {\cal G}_A (\td{q_1}) \bigg]~, \nn \\
\la \M{0}_A|\M{1}_A(\td{q_2})\ra &=& \gs^2 \, C_F \, \int_{\ell} \td{q_2} \bigg[
\frac{ |\M{0}_A |^2\, 4 q_2\cdot p_1}{(2M^2-2 q_2\cdot p_2)\,
 (s_{12}-2 q_2\cdot p_{12}+\imath 0)} +   {\cal G}_A (\td{q_2}) \bigg]~, \nn \\
\la \M{0}_A|\M{1}_A(\td{q_3})\ra &=& \gs^2 \, C_F \, \int_{\ell} \td{q_3} \bigg[
- \frac{ |\M{0}_A |^2\, 4 q_3\cdot p_2}{(2M^2+2 q_3\cdot p_1)\,(s_{12}+2 q_3\cdot p_{12})} 
+ {\cal G}_A (\td{q_3}) \bigg]~, \nn \\
\label{eq:alldual}
\eea 
where  $|\M{0}_A |^2$ are the Born squared amplitudes in \Eq{eq:bornsquared}, and the process dependent 
functions ${\cal G}_A (\td{q_i})$ for the process $\phi^*\to \qq$ are
\bea
{\cal G}_\phi (\td{q_1}) &=&  2 \, Y_q^2 \, C_A \,  \left( 1+\beta^2 + 2 c_q^2 \, 
\left( \frac{s_{12}}{2q_1\cdot p_1} - \frac{s_{12}}{2q_1\cdot p_2}\right) \right)~, \nn \\
{\cal G}_\phi (\td{q_2}) &=& 2 \, Y_q^2 \, C_A \,   \left( 
-\frac{m^2 \, s_{12} }{2M^2-2 q_2\cdot p_2} +
\frac{2(1-(2-\ep) \beta^2) \, s_{12} }{s_{12}-2 q_2\cdot p_{12}+\imath 0}  
- c_q^2 \frac{2(1-\ep) \, s_{12} }{s_{12}-2 q_2\cdot p_{12}+\imath 0}  \right)~, \nn \\
{\cal G}_\phi (\td{q_3}) &=& 2 \, Y_q^2 \, C_A \,  \left( 
-\frac{m^2 \, s_{12} }{2M^2+2 q_3\cdot p_1} +
\frac{2(1-(2-\ep) \beta^2) \, s_{12} }{s_{12}+2 q_3\cdot p_{12}}  
- c_q^2 \frac{2(1-\ep) \, s_{12} }{s_{12}+2 q_3\cdot p_{12}}  \right)~.
\label{eq:phidual}
\eea 
For $\gamma^*\to \qq$, they are
\bea
{\cal G}_\gamma (\td{q_1}) &=&  2\, (e\,  e_q)^2 \, C_A \,  \left( \left( 2 + \frac{m^2}{2(1-\ep)}\right)
\left( \frac{s_{12}}{2q_1\cdot p_1} - \frac{s_{12}}{2q_1\cdot p_2}\right) +2 \right)~, \nn \\
{\cal G}_\gamma (\td{q_2}) &=& 2\, (e\,  e_q)^2 \, C_A \,\left( 
\frac{m^2 \, s_{12} }{2(1-\ep) \, (2M^2-2 q_2\cdot p_2)} +
\frac{2\ep \, s_{12} + 4 q_2\cdot p_1}{s_{12}-2 q_2\cdot p_{12}+\imath 0}  \right)~, \nn \\
{\cal G}_\gamma (\td{q_3}) &=& 2\, (e\,  e_q)^2 \, C_A \, \left( 
\frac{m^2 \, s_{12} }{2(1-\ep) \, (2M^2+2 q_3\cdot p_1)} +
\frac{2\ep \, s_{12} - 4 q_3\cdot p_2}{s_{12}+2 q_3\cdot p_{12}}  \right)~.
\label{eq:gammadual}
\eea
Finally, for $Z^*\to \qq$, 
\bea
{\cal G}_Z (\td{q_1}) &=&  2\, C_A \, \frac{2(1-\ep)}{3-2\ep} \, \bigg[ g_{V,q}^2 \left( \left( 2 + \frac{m^2}{2(1-\ep)}\right)
\left( \frac{s_{12}}{2q_1\cdot p_1} - \frac{s_{12}}{2q_1\cdot p_2}\right) +2 \right) \nn \\
&+& g_{A,q}^2 \left( (1+\beta^2)
\left( \frac{s_{12}}{2q_1\cdot p_1} - \frac{s_{12}}{2q_1\cdot p_2} + 1 \right) 
- \frac{m^2}{2} \left( \frac{q_1\cdot p_2}{q_1\cdot p_1} - \frac{q_1\cdot p_1}{q_1\cdot p_2} \right) \right)\bigg]~, \nn \\
{\cal G}_Z (\td{q_2}) &=& 2\, C_A \, \frac{2(1-\ep)}{3-2\ep} \,  \bigg[ 
g_{V,q}^2 \, \left( 
\frac{m^2 \, s_{12} }{2(1-\ep) \, (2M^2-2 q_2\cdot p_2)} +
\frac{2\ep \, s_{12} + 4 q_2\cdot p_1}{s_{12}-2 q_2\cdot p_{12}+\imath 0}  \right) \nn \\
&+& g_{A,q}^2 \, \left( - \frac{m^2 \,(2 q_2\cdot p_{12}+s_{12})}{2 \, (2M^2-2 q_2\cdot p_2)} +
\frac{(m^2+2\ep\,\beta^2) \, s_{12} + 4q_2\cdot p_1 }{s_{12}-2 q_2\cdot p_{12}+\imath 0}  \right) \bigg]~, \nn \\
{\cal G}_Z (\td{q_3}) &=& 2\, C_A \, \frac{2(1-\ep)}{3-2\ep} \, \bigg[ 
g_{V,q}^2\, \left( \frac{m^2 \, s_{12} }{2(1-\ep) \, (2M^2+2 q_3\cdot p_1)} +
\frac{2\ep \, s_{12} -4 q_3\cdot p_2}{s_{12}+2 q_3\cdot p_{12}}  \right) \nn \\  
&+& g_{A,q}^2 \, \left( -\frac{m^2 \, (2 q_3\cdot p_{12}+s_{12})}{2 \, (2M^2+2 q_3\cdot p_1)} +
\frac{(m^2 + 2\ep \, \beta^2)\, s_{12} - 4 q_3\cdot p_2}{s_{12}+2 q_3\cdot p_{12}}  \right) \bigg]~.
\label{eq:Zdual}
\eea




\begin{thebibliography}{90}

\bibitem{Collins:1984xc}
  J.~C.~Collins,
  ``Renormalization : An Introduction to Renormalization, The Renormalization Group, and the Operator Product Expansion,''
 
\bibitem{Kinoshita:1962ur}
  T.~Kinoshita,
  ``Mass singularities of Feynman amplitudes,''
  J.\ Math.\ Phys.\  {\bf 3} (1962) 650.

\bibitem{Lee:1964is}
  T.~D.~Lee and M.~Nauenberg,
  ``Degenerate Systems and Mass Singularities,''
  Phys.\ Rev.\  {\bf 133} (1964) B1549.
  
  
\bibitem{Bollini:1972ui}
  C.~G.~Bollini and J.~J.~Giambiagi,
  ``Dimensional Renormalization: The Number of Dimensions as a Regularizing Parameter,''
  Nuovo Cim.\ B {\bf 12} (1972) 20.

\bibitem{'tHooft:1972fi}
  G.~'t Hooft and M.~J.~G.~Veltman,
  ``Regularization and Renormalization of Gauge Fields,''
  Nucl.\ Phys.\ B {\bf 44} (1972) 189.
  
  \bibitem{Cicuta:1972jf}
  G.~M.~Cicuta and E.~Montaldi,
  ``Analytic renormalization via continuous space dimension,''
  Lett.\ Nuovo Cim.\  {\bf 4} (1972) 329.
  
  \bibitem{Ashmore:1972uj}
  J.~F.~Ashmore,
  ``A Method of Gauge Invariant Regularization,''
  Lett.\ Nuovo Cim.\  {\bf 4} (1972) 289.
  
\bibitem{Kunszt:1992tn}
  Z.~Kunszt and D.~E.~Soper,
  ``Calculation of jet cross-sections in hadron collisions at order alpha-s**3,''
  Phys.\ Rev.\ D {\bf 46} (1992) 192.
 
\bibitem{Frixione:1995ms}
  S.~Frixione, Z.~Kunszt and A.~Signer,
  ``Three jet cross-sections to next-to-leading order,''
  Nucl.\ Phys.\ B {\bf 467} (1996) 399
  [hep-ph/9512328].

\bibitem{Catani:1996jh}
  S.~Catani and M.~H.~Seymour,
  ``The Dipole formalism for the calculation of QCD jet cross-sections at next-to-leading order,''
  Phys.\ Lett.\ B {\bf 378} (1996) 287
  [hep-ph/9602277].
        
\bibitem{Catani:1996vz}
  S.~Catani and M.~H.~Seymour,
  ``A General algorithm for calculating jet cross-sections in NLO QCD,''
  Nucl.\ Phys.\ B {\bf 485} (1997) 291
   [Nucl.\ Phys.\ B {\bf 510} (1998) 503]
  [hep-ph/9605323].

\bibitem{GehrmannDeRidder:2005cm}
  A.~Gehrmann-De Ridder, T.~Gehrmann and E.~W.~N.~Glover,
  ``Antenna subtraction at NNLO,''
  JHEP {\bf 0509} (2005) 056
  [hep-ph/0505111].
 
 \bibitem{Seth:2016hmv}
  S.~Seth and S.~Weinzierl,
  ``Numerical integration of subtraction terms,''
  Phys.\ Rev.\ D {\bf 93} (2016) 114031
  [arXiv:1605.06646 [hep-ph]].
 
\bibitem{Catani:2007vq}
  S.~Catani and M.~Grazzini,
  ``An NNLO subtraction formalism in hadron collisions and its application to Higgs boson production at the LHC,''
  Phys.\ Rev.\ Lett.\  {\bf 98} (2007) 222002
  [hep-ph/0703012].

\bibitem{Catani:2009sm}
  S.~Catani, L.~Cieri, G.~Ferrera, D.~de Florian and M.~Grazzini,
  ``Vector boson production at hadron colliders: a fully exclusive QCD calculation at NNLO,''
  Phys.\ Rev.\ Lett.\  {\bf 103} (2009) 082001
  [arXiv:0903.2120 [hep-ph]].
  
\bibitem{Czakon:2010td}
  M.~Czakon,
  ``A novel subtraction scheme for double-real radiation at NNLO,''
  Phys.\ Lett.\ B {\bf 693} (2010) 259
  [arXiv:1005.0274 [hep-ph]].

\bibitem{Bolzoni:2010bt}
  P.~Bolzoni, G.~Somogyi and Z.~Tr\'ocs\'anyi,
  ``A subtraction scheme for computing QCD jet cross sections at NNLO: integrating the iterated singly-unresolved subtraction terms,''
  JHEP {\bf 1101} (2011) 059
  [arXiv:1011.1909 [hep-ph]].
  
\bibitem{DelDuca:2015zqa}
  V.~Del Duca, C.~Duhr, G.~Somogyi, F.~Tramontano and Z.~Tr\'ocs\'anyi,
  ``Higgs boson decay into b-quarks at NNLO accuracy,''
  JHEP {\bf 1504} (2015) 036
  [arXiv:1501.07226 [hep-ph]].

\bibitem{Boughezal:2015dva}
  R.~Boughezal, C.~Focke, X.~Liu and F.~Petriello,
  ``$W$-boson production in association with a jet at next-to-next-to-leading order in perturbative QCD,''
  Phys.\ Rev.\ Lett.\  {\bf 115} (2015) 062002
  [arXiv:1504.02131 [hep-ph]].

\bibitem{Gaunt:2015pea}
  J.~Gaunt, M.~Stahlhofen, F.~J.~Tackmann and J.~R.~Walsh,
  ``N-jettiness Subtractions for NNLO QCD Calculations,''
  JHEP {\bf 1509} (2015) 058
  [arXiv:1505.04794 [hep-ph]].
 
 \bibitem{DelDuca:2016ily}
  V.~Del Duca, C.~Duhr, A.~Kardos, G.~Somogyi, Z.~Szor, Z.~Tr\'ocs\'anyi and Z.~Tulip\'ant,
  ``Jet production in the CoLoRFulNNLO method: event shapes in electron-positron collisions,''
  arXiv:1606.03453 [hep-ph].

 \bibitem{DelDuca:2016csb}
  V.~Del Duca, C.~Duhr, A.~Kardos, G.~Somogyi and Z.~Tr\'ocs\'anyi,
  ``Three-jet production in electron-positron collisions using the CoLoRFulNNLO method,''
  arXiv:1603.08927 [hep-ph].

\bibitem{Catani:2002hc}
  S.~Catani, S.~Dittmaier, M.~H.~Seymour and Z.~Trocsanyi,
  ``The Dipole formalism for next-to-leading order QCD calculations with massive partons,''
  Nucl.\ Phys.\ B {\bf 627} (2002) 189
  [hep-ph/0201036].

\bibitem{GehrmannDeRidder:2009fz}
  A.~Gehrmann-De Ridder and M.~Ritzmann,
  ``NLO Antenna Subtraction with Massive Fermions,''
  JHEP {\bf 0907} (2009) 041
  [arXiv:0904.3297 [hep-ph]].

\bibitem{Abelof:2012he}
  G.~Abelof, O.~Dekkers and A.~Gehrmann-De Ridder,
  ``Antenna subtraction with massive fermions at NNLO: Double real initial-final configurations,''
  JHEP {\bf 1212} (2012) 107
  [arXiv:1210.5059 [hep-ph]].
  
\bibitem{Abelof:2014fza}
  G.~Abelof, A.~Gehrmann-De Ridder, P.~Maierhofer and S.~Pozzorini,
  ``NNLO QCD subtraction for top-antitop production in the $ q\overline{q} $ channel,''
  JHEP {\bf 1408} (2014) 035
  [arXiv:1404.6493 [hep-ph]].

\bibitem{Bonciani:2015sha}
  R.~Bonciani, S.~Catani, M.~Grazzini, H.~Sargsyan and A.~Torre,
  ``The $q_T$ subtraction method for top quark production at hadron colliders,''
  Eur.\ Phys.\ J.\ C {\bf 75} (2015) no.12,  581
  [arXiv:1508.03585 [hep-ph]].

\bibitem{Catani:2008xa}
  S.~Catani, T.~Gleisberg, F.~Krauss, G.~Rodrigo and J.~C.~Winter,
  ``From loops to trees by-passing Feynman's theorem,''
  JHEP {\bf 0809} (2008) 065
  [arXiv:0804.3170 [hep-ph]].

\bibitem{Rodrigo:2008fp}
  G.~Rodrigo, S.~Catani, T.~Gleisberg, F.~Krauss and J.~C.~Winter,
  ``From multileg loops to trees (by-passing Feynman's Tree Theorem),''
  Nucl.\ Phys.\ Proc.\ Suppl.\  {\bf 183} (2008) 262
  [arXiv:0807.0531 [hep-th]].


\bibitem{Bierenbaum:2010cy}
  I.~Bierenbaum, S.~Catani, P.~Draggiotis and G.~Rodrigo,
  ``A Tree-Loop Duality Relation at Two Loops and Beyond,''
  JHEP {\bf 1010} (2010) 073
  [arXiv:1007.0194 [hep-ph]].

\bibitem{Bierenbaum:2012th}
  I.~Bierenbaum, S.~Buchta, P.~Draggiotis, I.~Malamos and G.~Rodrigo,
  ``Tree-Loop Duality Relation beyond simple poles,''
  JHEP {\bf 1303} (2013) 025
  [arXiv:1211.5048 [hep-ph]].


\bibitem{Buchta:2014dfa}
  S.~Buchta, G.~Chachamis, P.~Draggiotis, I.~Malamos and G.~Rodrigo,
  ``On the singular behaviour of scattering amplitudes in quantum field theory,''
  JHEP {\bf 1411} (2014) 014
  [arXiv:1405.7850 [hep-ph]].

\bibitem{Buchta:2015xda}
  S.~Buchta,
  ``Theoretical foundations and applications of the Loop-Tree Duality in Quantum Field Theories,'' PhD thesis, Universitat de Val\`encia, 2015,
  arXiv:1509.07167 [hep-ph].
  
\bibitem{Buchta:2015wna}
  S.~Buchta, G.~Chachamis, P.~Draggiotis and G.~Rodrigo,
  ``Numerical implementation of the Loop-Tree Duality method,''
  arXiv:1510.00187 [hep-ph].

\bibitem{Buchta:2015jea}
  S.~Buchta, G.~Chachamis, P.~Draggiotis, I.~Malamos and G.~Rodrigo,
  ``Towards a Numerical Implementation of the Loop-Tree Duality Method,''
  Nucl.\ Part.\ Phys.\ Proc.\  {\bf 258-259} (2015) 33
  [arXiv:1509.07386 [hep-ph]].

\bibitem{Hernandez-Pinto:2015ysa}
  R.~J.~Hern\' andez-Pinto, G.~F.~R.~Sborlini and G.~Rodrigo,
  ``Towards gauge theories in four dimensions,''
  JHEP {\bf 1602} (2016) 044
  [arXiv:1506.04617 [hep-ph]].

\bibitem{Sborlini:2015uia}
  G.~F.~R.~Sborlini, R.~Hern\'andez-Pinto and G.~Rodrigo,
  ``From dimensional regularization to NLO computations in four dimensions,''
  PoS EPS-HEP {\bf 2015} (2015) 479
  [arXiv:1510.01079 [hep-ph]].

\bibitem{Sborlini:2016fcj}
  G.~F.~R.~Sborlini,
  ``Loop-tree duality and quantum field theory in four dimensions,''
  PoS RADCOR {\bf 2015} (2015) 082 
  [arXiv:1601.04634 [hep-ph]].
  
\bibitem{Sborlini:2016gbr}
  G.~F.~R.~Sborlini, F.~Driencourt-Mangin, R.~Hern\'andez-Pinto and G.~Rodrigo,
  ``Four-dimensional unsubtraction from the loop-tree duality,''
  arXiv:1604.06699 [hep-ph].
 
\bibitem{ll2016}
  G.~Rodrigo, F.~Driencourt-Mangin, G.~F.~R.~Sborlini and R.~Hern\'andez-Pinto,
  ``Applications of the loop-tree duality,''
  PoS LL {\bf 2016} (2016) 037.
 
\bibitem{Soper:1998ye}
  D.~E.~Soper,
  ``QCD calculations by numerical integration,''
  Phys.\ Rev.\ Lett.\  {\bf 81} (1998) 2638
  [hep-ph/9804454].

\bibitem{Soper:1999xk}
  D.~E.~Soper,
  ``Techniques for QCD calculations by numerical integration,''
  Phys.\ Rev.\ D {\bf 62} (2000) 014009
  [hep-ph/9910292].

\bibitem{Soper:2001hu}
  D.~E.~Soper,
  ``Choosing integration points for QCD calculations by numerical integration,''
  Phys.\ Rev.\ D {\bf 64} (2001) 034018
  [hep-ph/0103262].

\bibitem{Kramer:2002cd}
  M.~Kramer and D.~E.~Soper,
  ``Next-to-leading order numerical calculations in Coulomb gauge,''
  Phys.\ Rev.\ D {\bf 66} (2002) 054017
  [hep-ph/0204113].

\bibitem{Becker:2010ng}
  S.~Becker, C.~Reuschle and S.~Weinzierl,
  ``Numerical NLO QCD calculations,''
  JHEP {\bf 1012} (2010) 013
  [arXiv:1010.4187 [hep-ph]].

\bibitem{Becker:2012aqa}
  S.~Becker, C.~Reuschle and S.~Weinzierl,
  ``Efficiency Improvements for the Numerical Computation of NLO Corrections,''
  JHEP {\bf 1207} (2012) 090
  [arXiv:1205.2096 [hep-ph]].
  
\bibitem{Pittau:2012zd}
  R.~Pittau,
  ``A four-dimensional approach to quantum field theories,''
  JHEP {\bf 1211} (2012) 151
  [arXiv:1208.5457 [hep-ph]].
  
\bibitem{Donati:2013iya}
  A.~M.~Donati and R.~Pittau,
  ``Gauge invariance at work in FDR: $H \to \gamma \gamma$,''
  JHEP {\bf 1304} (2013) 167
  [arXiv:1302.5668 [hep-ph]].
  
\bibitem{Fazio:2014xea}
  R.~A.~Fazio, P.~Mastrolia, E.~Mirabella and W.~J.~Torres Bobadilla,
  ``On the Four-Dimensional Formulation of Dimensionally Regulated Amplitudes,''
  Eur.\ Phys.\ J.\ C {\bf 74} (2014) no.12,  3197
  [arXiv:1404.4783 [hep-ph]].
  
\bibitem{Feynman:1963ax}
  R.~P.~Feynman,
  ``Quantum theory of gravitation,''
  Acta Phys.\ Polon.\  {\bf 24} (1963) 697.

\bibitem{Feynman:1972mt}
  R.~P.~Feynman,
  ``Closed Loop And Tree Diagrams. (talk),''
  In *Brown, L.M. (ed.): Selected papers of Richard Feynman* 867-887

\bibitem{Ellis:2007qk}
  R.~K.~Ellis and G.~Zanderighi,
  ``Scalar one-loop integrals for QCD,''
  JHEP {\bf 0802} (2008) 002 
  [arXiv:0712.1851 [hep-ph]].
  
\bibitem{Rodrigo:1999qg}
  G.~Rodrigo, M.~S.~Bilenky and A.~Santamaria,
  ``Quark mass effects for jet production in e+ e- collisions at the next-to-leading order: Results and applications,''
  Nucl.\ Phys.\ B {\bf 554} (1999) 257
  [hep-ph/9905276].
  
\bibitem{Bilenky:1994ad}
  M.~S.~Bilenky, G.~Rodrigo and A.~Santamaria,
  ``Three jet production at LEP and the bottom quark mass,''
  Nucl.\ Phys.\ B {\bf 439} (1995) 505
  [hep-ph/9410258].




\end{thebibliography}
\end{document}